\documentclass[journal,twoside]{IEEEtran}
\usepackage{multirow}
\usepackage[T1]{fontenc}
\usepackage{booktabs}
\usepackage{cite}
\usepackage{textcase}
\usepackage{amsmath,amssymb,amsfonts}
\usepackage{array}
\usepackage{float}
\usepackage{graphicx}
\usepackage{mathabx}
\usepackage{orcidlink}
\hypersetup{hidelinks}
\usepackage{textcomp}
\usepackage{lettrine}
\usepackage{comment}
\usepackage{xcolor}
\usepackage{pgfplotstable}
\usepackage{pgfplots}
\usepackage{siunitx} 
\usepackage{textgreek} 
\usepackage{ragged2e}
\usepackage{hyperref}
\usepackage[ruled,vlined]{algorithm2e}
\usepgfplotslibrary{dateplot}
\pgfplotsset{compat=newest}
\usepackage{filecontents}
\usepackage{threeparttable}
\usepackage{tabularx}
\newcolumntype{Y}{>{\centering\arraybackslash}X}
\def\BibTeX{{\rm B\kern-.05em{\sc i\kern-.025em b}\kern-.08em
    T\kern-.1667em\lower.7ex\hbox{E}\kern-.1
    25emX}}
\definecolor{colour1}{HTML}{4477AA} % blue
\definecolor{colour2}{HTML}{F748A5} % pink
\definecolor{colour3}{HTML}{CCBB44} % yellow
\usepackage[caption=false,font=normalsize,labelfont=sf,textfont=sf]{subfig}
\usepackage{stfloats}
\usepackage{url}
\usepackage{verbatim}
\def\BibTeX{{\rm B\kern-.05em{\sc i\kern-.025em b}\kern-.08em
    T\kern-.1667em\lower.7ex\hbox{E}\kern-.125emX}}
\usepackage{balance}

\newcommand*\circled[1]{\protect\tikz[baseline=(char.base)]{\protect\node[shape=circle,fill=black!10,inner sep=1.2pt] (char) {\textcolor{black} #1};}}  

\newcommand*\circledwhite[1]{\protect\tikz[baseline=(char.base)]{\protect\node[shape=circle,fill=black,inner sep=1.2pt] (char) {\textcolor{white}{#1}};}}

\usepackage{tablefootnote}

\begin{document}

\title{Acore-CIM: build accurate and reliable mixed-signal CIM cores with RISC-V controlled self-calibration}

\author{Omar Numan\textsuperscript{*}\textsuperscript{\orcidlink{0009-0004-5679-8709}}, \IEEEmembership{Student Member}, Gaurav Singh\textsuperscript{*}\textsuperscript{\orcidlink{0009-0009-7801-4721}},
\IEEEmembership{Student Member}, Kazybek Adam,
\IEEEmembership{Student Member}, Jelin Leslin\textsuperscript{\orcidlink{0000-0002-6963-6192}},
\IEEEmembership{Student Member}, Aleksi Korsman\textsuperscript{\orcidlink{0009-0000-4683-0884}}, \IEEEmembership{Student Member}, Otto Simola, \IEEEmembership{Student Member}, Marko Kosunen\textsuperscript{\orcidlink{0000-0002-2723-1859}}, \IEEEmembership{Member, IEEE}, Jussi Ryynänen\textsuperscript{\orcidlink{0000-0002-2241-092X}}, \IEEEmembership{Member, IEEE}, and Martin Andraud\textsuperscript{\orcidlink{0000-0002-8829-1054}}, \IEEEmembership{Member, IEEE}

\thanks{\textsuperscript{*}These authors contributed equally to this work.}
\thanks{All authors are with the Department of Electronics and Nanoengineering, Aalto University, 02150 Espoo, Finland (e-mail: firstname.lastname@aalto.fi).}
\thanks{Martin Andraud and Gaurav Singh are also with ICTEAM, UCLouvain, Belgium (e-mail: firstname.lastname@uclouvain.be).}
\thanks{This work is funded by Academy of Finland projects: EHIR (grant 13334487) and WHISTLE (grant 332218).}
}

\maketitle

\begin{abstract}
Developing accurate and reliable Compute-In-Memory (CIM) architectures is becoming a key research focus to accelerate Artificial Intelligence (AI) tasks on hardware, particularly Deep Neural Networks (DNNs). In that regard, there has been significant interest in analog and mixed-signal CIM architectures aimed at increasing the efficiency of data storage and computation to handle the massive amount of data needed by DNNs. Specifically, resistive mixed-signal CIM cores are pushed by recent progresses in emerging Non-Volatile Memory (eNVM) solutions. Yet, mixed-signal CIM computing cores still face several integration and reliability challenges that hinder their large-scale adoption into end-to-end AI computing systems. In terms of integration, resistive and eNVM-based CIM cores need to be integrated with a control processor to realize end-to-end AI acceleration. Moreover, SRAM-based CIM architectures are still more efficient and easier to program than their eNVM counterparts. In terms of reliability, analog circuits are more susceptible to variations, leading to computation errors and degraded accuracy. This work addresses these two challenges by proposing a self-calibrated mixed-signal CIM accelerator SoC, fabricated in 22-nm FDSOI technology. The integration is facilitated by (1) the CIM architecture, combining the density and ease of SRAM-based weight storage with multi-bit computation using linear resistors, and (2) an open-source programming and testing strategy for CIM systems. The accuracy and reliability are enabled through an automated RISC-V controlled on-chip calibration, allowing us to improve the compute SNR by 25 to 45\% across multiple columns to reach 18-24 dB. To showcase further integration possibilities, we show how our proof-of-concept SoC can be extended to recent high-density linear resistor technologies for enhanced computing performance.  
\end{abstract}

\begin{IEEEkeywords}
AI accelerator, computing-in-memory (CIM), deep neural network (DNN), self-calibration, RISC-V.
\end{IEEEkeywords}

\section{Introduction}
\lettrine{T}{he} recent push for embedded Artificial Intelligence (AI), particularly Deep Neural Networks (DNNs), in resource-limited embedded systems has raised a critical need for energy-efficient and dedicated AI processors, or \emph{accelerators}. In this context, Compute-In-Memory (CIM) cores have received significant attention \cite{SSMHSALHY22}. CIM efficiently exploits the inherent parallelism in DNN computations, which predominantly involve large numbers of Multiply-And-Accumulate (MAC) operations, by performing MAC directly in memory while drastically reducing data transfers. The energy efficiency can be greatly improved using mixed-signal CIM, based on analog computing principles \cite{Mur21}. Here, MAC computations follow Ohm's law, with DNN weights encoded either as variable resistances (conductances) \cite{PLWLCL24} or as capacitances \cite{JYSS20}, and multiplied by analog inputs. Accumulation occurs natively with currents \cite{PLWLCL24, CLLHL18} or through charge sharing \cite{JYSS20, VRNV19}. Focusing on resistive-based CIM cores, advancements are closely associated with the development of emerging Non-Volatile Memories (eNVMs), aiming to provide multi-bit weight encoding, easy access, and non-volatility. Emerging NVMs include Resistive RAM (RRAM) \cite{HYFHJ22,HWHHS22,PGKRG22,YCKCCR22}, Spin-Transfer Torque Magnetoresistive RAM (STT-MRAM) \cite{AMSOT22}, Ferroelectric RAM (FeRAM) \cite{SKKRB21}, and Phase Change Memory (PCM) \cite{gallo22}. Despite the potential of these emerging technologies, numerous challenges remain in the development of reliable mixed-signal CIM solutions. In particular, due to the analog nature of computations, mixed-signal CIM designs still need to overcome challenges to be used at a larger scale. 

As a starting point, CIM architectures must ensure sufficient computational accuracy for typical DNN tasks. In particular, a compute Signal-to-Noise Ratio (SNR) in the 20–22 dB range limits the computational accuracy loss to 5\% compared to fixed-point computation \cite{SaSh18}. In this work, we adopt the compute SNR definition from \cite{Shan22}, which explicitly accounts for both noise and distortion. Focusing on resistive-based CIM arrays, a recent evaluation carried out with multiple eNVM technologies \cite{RoSh24} pointed out that this cannot be achieved without calibration, i.e., the inherent variability of the devices does not guarantee sufficient compute SNR margin. This susceptibility led to research on calibration strategies for various CIM cores \cite{PLWLCL24,SENAS21,SSC24,CJSLZ22,HWHHS22} and resistive-based architectures \cite{Li22,YaXiYi22}. However, calibration techniques typically include area, power, or timing overheads, limiting the system's efficiency. Hence, it is necessary to develop automated and integrated calibration routines that minimize system-level overheads in area, power, and latency while achieving the required compute SNR. 

Based on this initial observation, we identify three key challenges for developing more accurate and reliable resistive-based CIM cores: \textbf{\circled{1}} providing efficient and reliable computation units, still benefiting from SRAM integration and weight storage but opening the path for eNVMs, particularly high-density linear resistive technologies; \textbf{\circled{2}} simplifying the integration of CIM cores in complete AI acceleration systems with RISC-V processors through open-source programming, testing and calibration routines, and \textbf{\circled{3}} offering automated self-calibration solutions, through autonomous RISC-V controlled self-calibration to achieve the required accuracy requirements. In this work, we propose to take a holistic view on tackling these challenges towards accurate and reliable mixed-signal CIM designs, specifically:  

To tackle \textbf{Ch.\circled{1}}, we propose to design a \emph{reliability-focused} MAC cell that employs a multiplication scheme based on a multi-bit current-mode R-2R Multiplying Digital-to-Analog Converter (MDAC) topology, where each weight bit is efficiently stored in a 6T-SRAM cell \cite{KiHaBe04,LaHaAy11}. A key advantage of this approach is the use of SRAM weight storage and more reliable \emph{linear} resistors instead of multi-state eNVM devices, for instance High Density Linear Resistor (HDLR) technologies \cite{SKKRB21,CKHLLH21}. HDLRs have a higher value and smaller area than typical poly resistors in CMOS, and can be post-processed on top of CMOS wafers. To tackle \textbf{Ch.\circled{2}}, we design a complete \emph{proof-of-concept} SoC composed of a CIM core and a RISC-V control processor. To ease integration, our programming framework is fully open-source, promoting the test and characterization of CIM-based designs in full systems. 

To tackle \textbf{Ch.\circled{3}}, we demonstrate on our SoC an automated Built-In Self-Calibration (BISC) routine, operating in real-time and fully controlled by the RISC-V core. The BISC mitigates offset and gain errors, improving the compute SNR by 25 to 45\% (6-8 dB) in all CIM columns to push the compute SNR to 18--24 dB. This ensures sufficient computational accuracy while minimizing the calibration overheads by making use of the existing AI computation system.  
The proof-of-concept SoC has been fabricated and measured in a 22-nm FD-SOI technology. In addition, all hardware/software integration and calibration routines have been released as open-source, including a model of the complete SoC, allowing thorough testing of the architecture and programming routines in software before deployment to the physical device.

This paper is organized as follows: Section \ref{sec:SOTA} reviews challenges in design and reliability of mixed-signal CIM, focusing on resistive-based arrays; Section \ref{sec:System} details our proof-of-concept accelerator SoC use as a baseline for the work; sections \ref{sec:CIM}, \ref{sec:git}, and \ref{sec:BISC} explain how the three aforementioned challenges are tackled in our SoC architecture; section \ref{sec:measurements} presents measurements results of the proof-of-concept chip, and section \ref{sec:Concl} concludes the work.

\section{Design and Reliability Challenges of Mixed-Signal CIM}\label{sec:SOTA}

\subsection{SRAM versus eNVM CIM design}\label{sec:SRAMeNVM}

On the one hand, SRAM-based CIM has evolved from binary and ternary content-addressable memory for search applications to architectures supporting logic and MAC operations, particularly in AI edge devices, driven by its high endurance, fast write speeds, low write energy, and compatibility with modern logic processes \cite{JXHCC21}. These systems use various cell architectures: compact 6T cells for energy-efficient MAC operations or less area-efficient 7T, 8T, and 10T cells that improve signal margin, reduce read disturbances, and enable more complex functions \cite{JXHCC21,ShRo22}. Hence, SRAM-based CIM systems now support multi-bit MAC operations \cite{PLWLCL24}. Despite these advances, challenges remain; for instance, increasing MAC precision with compact 6T-SRAM cells typically reduces signal margins, which impacts readout accuracy and increases the likelihood of read disturbances, especially when the signal margin falls below the ADC input offset due to limited bit-line voltage swings. Increasing to multi-bit MAC requires additional cycles, which increases latency and reduces area efficiency. In addition, activating multiple word-lines improves energy efficiency but can cause read disturbances due to expanded bit-line voltage swings, potentially leading to data flipping. Finally, process variations cause input offsets in analog readout circuits, requiring larger signal margins to maintain precision.

On the other hand, eNVM-based CIM cores combine memory and processing in a single cell. Recent developments in these architectures \cite{HYFHJ22,HWHHS22,PGKRG22,YCKCCR22} have utilized metal-oxide layers for high-speed switching and computation. For instance, RRAMs employ a metal-insulator-metal structure where conductive filaments form and rupture via redox reactions, making them ideal for high-density, low-power, multi-bit memory applications. By enabling both logic operations and data storage within the same cell, MAC operations can be performed directly in memory, thereby reducing data movement and significantly enhancing energy efficiency and speed.

However, several challenges still affect the design of eNVM-based systems \cite{LUSRK24,YCKCCR22,SCKCBA24,SBDJHH22}. For instance, eNVM cells may struggle with higher multi-bit MAC operations. Solutions using an 8-to-1 bit-line multiplexer and serial access reduce area and energy efficiencies by up to 97\% and 94\%, respectively, compared to 1-bit operations \cite{HWHHS22}. Also, eNVM cells may require precise high-voltage pulses for programming, adding complexity to the circuit, while additional peripherals may be necessary to avoid unintentional overwriting (write-disturb issues). In terms of computation, increasing the number of active word-lines reduces the current-domain sensing margin, leading to errors from parasitic resistances in bit-lines and multiplexers. Finally, analog non-idealities (spec. in the summing amplifier) impact the linearity of current sensing, requiring complex corrective measures.

In that regard, SRAM-based CIM cores remain up to 7× more efficient than their eNVM-based counterparts \cite{ShRo22}. This can be explained as \textbf{(I)} SRAMs continue to benefit significantly from technology scaling, \textbf{(II)} eNVM are still far from the level of integration of SRAMs and typically harder to program, and \textbf{(III)} the use of analog weights can impose higher demands on peripherals (amplifiers and data converters), thus reducing area and energy efficiency. As a result, it remains beneficial to find a MAC computing unit that can benefit at the same time from efficient programming and weight storage with SRAM, while opening the possibility of integrating eNVMs and HDLRs for power-efficient computation.

\subsection{CIM integration through control processors}

To build a complete AI system, a mixed-signal CIM core must be integrated with a control processor that manages DNN tasks and handles all non-MAC operations. This is typically achieved using a RISC-V processor capable of managing one or multiple CIM cores \cite{SSMHSALHY22,DIANA22}. 

\subsection{Reliability of mixed-signal CIM cores}

\begin{figure}[t]
    \centering
    \includegraphics[width=0.9\columnwidth]{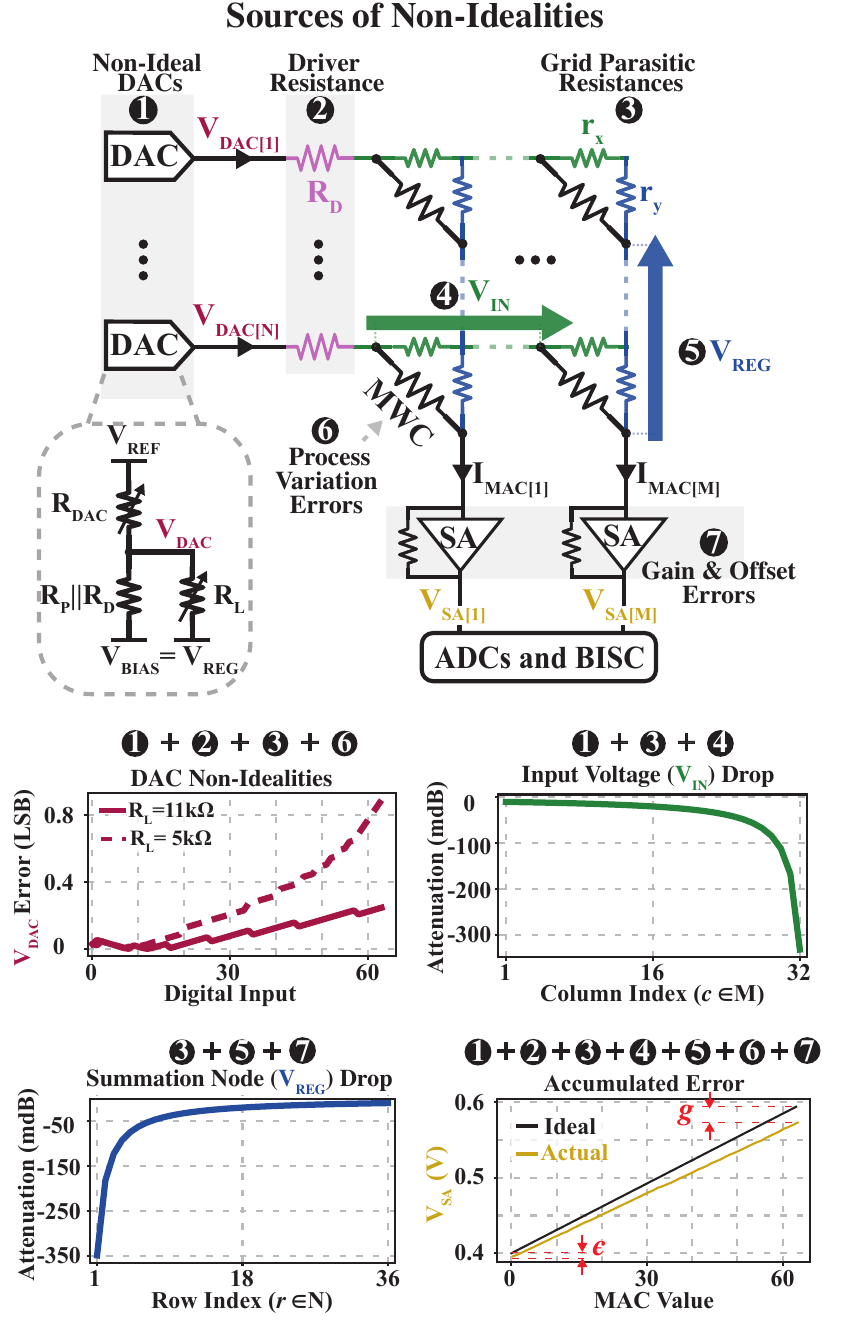}
    \caption{Key sources of non-idealities in resistive-based CIM cores: DAC output errors, column-wise input attenuation, and row-wise summation node voltage drops. The MAC error plot highlights discrepancies due to gain ($g$) and offset ($\epsilon$) deviations.}
    \label{fig:Challenges}
\end{figure}

Resistive-based CIM cores are inherently vulnerable to various non-idealities that degrade computational accuracy. These arise from peripheral components, interconnect resistances, and device-level variations, leading to errors that propagate across the CIM core \cite{RxNN}. Maintaining signal integrity for DNN workloads requires a thorough understanding of these effects and the implementation of corrective measures.

Fig. \ref{fig:Challenges} provides a detailed breakdown of the primary non-idealities affecting resistive-based CIM cores, categorized as follows:

\paragraph{Non-ideal DACs \textbf{\circledwhite{1}}} Input DACs exhibit inaccuracies due to finite output impedance, load dependency, and process variations, causing deviations in applied voltages across CIM array rows.
\paragraph{Driver resistance \textbf{\circledwhite{2}}} The finite output resistance of driver circuits ($R_D$) introduces voltage drops along the rows, further altering applied input levels.
\paragraph{Parasitic wire resistances \textbf{\circledwhite{3}}} Interconnect resistances ($r_x, r_y$) introduce voltage attenuation across both rows and columns, leading to signal degradation.
\paragraph{Input signal attenuation \textbf{\circledwhite{4}}} The combined effects of driver resistance and wire parasitics progressively reduce the input voltage ($V_{IN}$) across columns.
\paragraph{Summation node regulation voltage drop \textbf{\circledwhite{5}}} Parasitic resistances along the columns affect the virtual ground regulation voltage ($V_{REG}$), reducing accuracy in the accumulated current summation.
\paragraph{MAC cell conductance variability \textbf{\circledwhite{6}}} After chip fabrication, device-level variations and mismatch can shift operating points and add additional errors.
\paragraph{Summing amplifier offset and gain errors \textbf{\circledwhite{7}}} The Summing Amplifier (SA) at the column outputs introduces offset and gain errors—input-referred offset voltage shifts MAC outputs, causing systematic bias, while a finite open-loop gain reduces I-V conversion accuracy. Additionally, the finite input impedance disrupts the virtual ground ($V_{REG}$), altering the accumulated currents. These distortions collectively degrade computational precision and linearity.

The plots in Fig. \ref{fig:Challenges} illustrate how these non-idealities affect CIM operation:

\begin{itemize}
\item DAC Non-Idealities (\textbf{\circledwhite{1}+\circledwhite{2}+\circledwhite{3}+\circledwhite{6}})$\rightarrow$Increased DAC output ($V_{DAC}$) errors due to load resistance ($R_L$) variations.
\item Input Voltage Drop (\textbf{\circledwhite{1}+\circledwhite{3}+\circledwhite{4}})$\rightarrow$Progressive attenuation of $V_{IN}$ across columns, highlighting the effect of interconnect resistances.
\item Summation Node Voltage Drop (\textbf{\circledwhite{3}+\circledwhite{5}+\circledwhite{7}})$\rightarrow$Decrease in $V_{REF}$ across rows due to parasitic resistance buildup.
\item Accumulated Error (\textbf{\circledwhite{1} to \circledwhite{7}})$\rightarrow$Discrepancies between ideal and actual $V_{SA}$, where gain ($g$) and offset ($\epsilon$) errors emerge from the cumulative impact of these non-idealities. Error accumulation in CIM arrays is not strictly correlated with physical distance; it depends on the array state, programmed conductances, and input patterns. In addition to systematic errors, random variations—such as thermal noise, flicker noise, and inherent device mismatches—also contribute to performance variability.
\end{itemize}

In summary, resistive-based CIM cores suffer from interdependent non-idealities, where errors vary with array size, conductance states, input patterns, process variations, and random noise sources. To ensure computational accuracy and reliability, a robust self-calibration mechanism is essential. Section \ref{sec:BISC} details our BISC-based calibration strategy, which primarily targets systematic errors while recognizing that a residual random error floor remains.

\section{Proof-of-Concept AI Accelerator SoC}\label{sec:System}

\begin{figure*}[t]
    \centering
    \includegraphics[width=1.8\columnwidth]{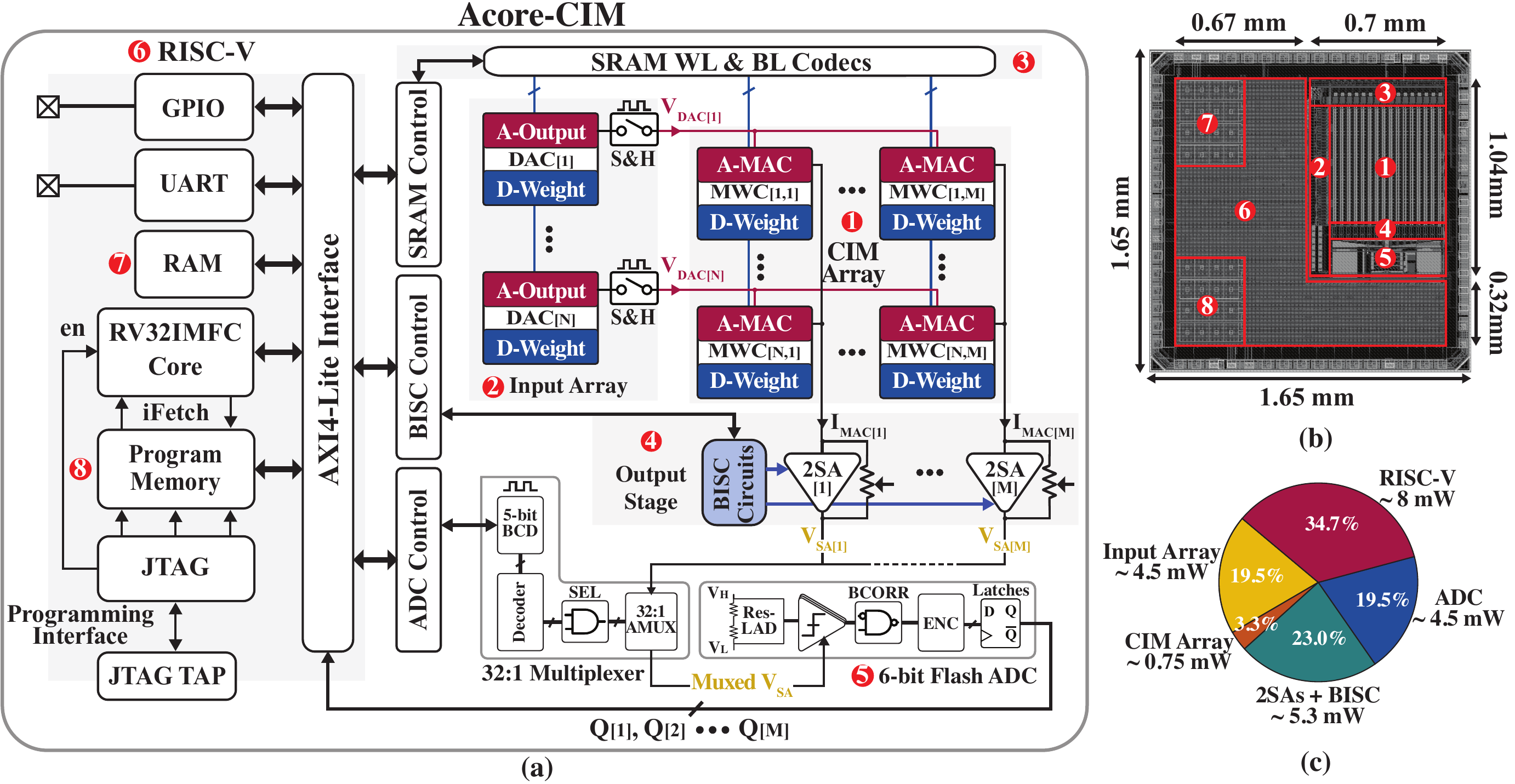}
    %\vspace{-5pt}
    \caption{Proof-of-concept Acore-CIM SoC composed of a 32-bit RISC-V core controlling a $36\times32$ mixed-signal CIM core through AXI4-Lite interface. (a) Detailed block diagram; b) Die microphotograph; (c) Power distribution of the SoC prototype.}
    %\vspace{-7pt}
    \label{fig:CHIP_BD}
\end{figure*}

\noindent Fig. \ref{fig:CHIP_BD}(a) depicts our Acore-CIM SoC, comprising a RISC-V core, a mixed-signal CIM accelerator, and an AXI4-Lite interface for interconnecting the components. This section provides a baseline of the overall architecture and its different blocks. 

\subsection{RISC-V core and AXI4-Lite interface}\label{sec:risc-v}
The backbone of the proposed A-core SoC is a custom-designed, 5-stage RISC-V compliant RV32IMFC processor core, illustrated in Fig. \ref{fig:CHIP_BD}(a) and made fully open-source \cite{ACore_2024}. It supports integer multiplication and division, single-precision floating-point operations, and compressed 16-bit instructions through the M, F, and C extensions. The core, along with the AXI4-Lite interface, was developed using Chisel HDL \cite{Chisel_2024}. It is fully parameterized, allowing selection of the RISC-V extensions to suit specific applications. In this project, all extensions were enabled, with the core achieving benchmark scores of 0.628 DMIPS/MHz and 2.07 Coremark/MHz.

The core is programmed over a JTAG interface. Communication between the processor and the CIM core is handled through a data bus following the AXI4-Lite protocol, a simplified subset of the AXI4 protocol. AXI4-Lite limits the data width to 32 or 64 bits and removes burst capability. The core data bus width is set to 32 bits, allowing 32-bit transfers per clock cycle under optimal conditions. Peripherals such as UART and GPIO are also connected to the interconnect, providing processor-controllable interfaces for external communication. The CIM core contains control registers, clocked in the RISC-V core clock domain, interfaced via AXI4-Lite. This processor-programmable control interface is, for instance, used to implement RISC-V controlled calibration.

\subsection{CIM accelerator core}\label{sec:CIM-core}

\begin{figure}[t]
    \centering
    \includegraphics[width=\columnwidth]{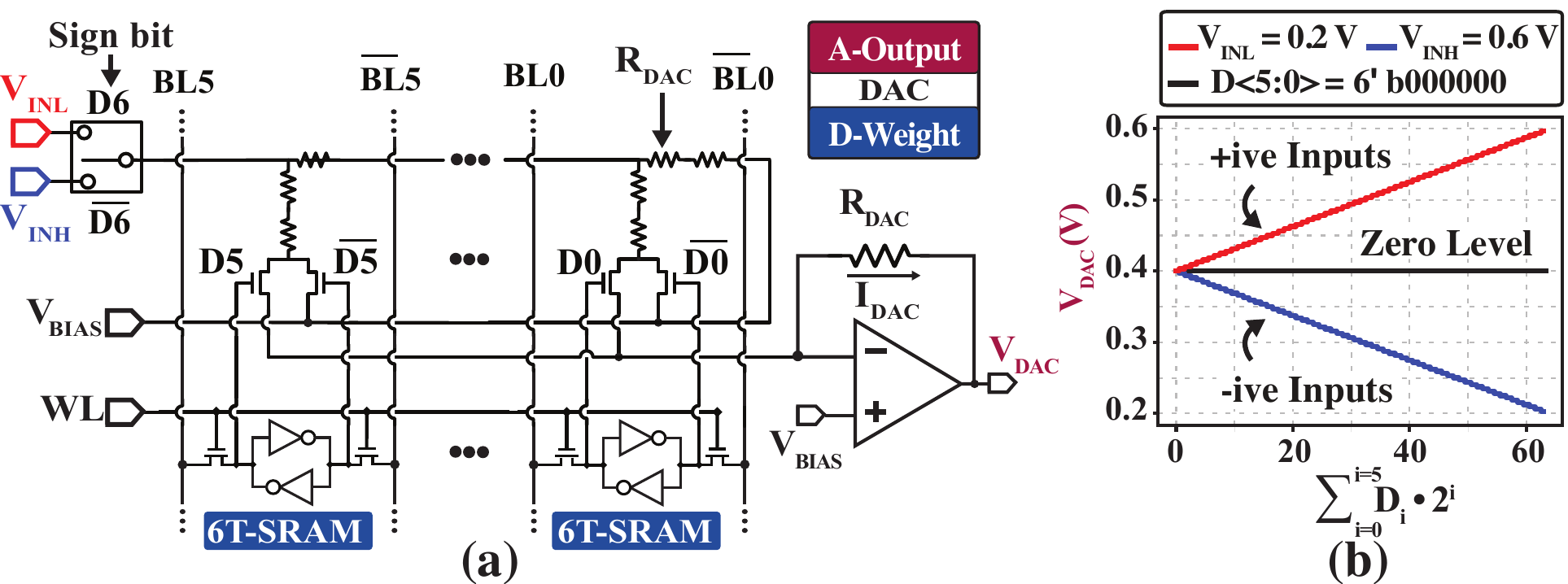}
    \caption{(a) 6-bit R-2R MDAC Input DAC Cell with an extra sign bit for dual-polarity operation; (b) Transfer characteristics for signed inputs.}
    \label{fig:IPDAC}
\end{figure}

The CIM core comprises 4 main blocks:
\subsubsection{Input array} The input array consists of a DAC chain, where an SRAM-based decoder programs the $N$ resistive DACs setting their digital input values. These DACs convert stored digital values into analog signals, which are then buffered by $N$ sample-and-hold (S\&H) circuits to drive signals across the $M$ columns. The S\&H circuits ensure signal integrity, stability, and precise timing, stabilizing the analog output of the DACs, mitigating noise, and synchronizing inputs before entering the CIM array. These circuits add design flexibility, enabling the accelerator to handle both analog and digital inputs with minimal architectural changes. Fig. \ref{fig:IPDAC}(a) illustrates the input DAC cell, which employs an R-2R MDAC topology with a precision of B\textsubscript{D} = 6+1 bits, i.e., the input consists of a 6-bit digital value ($D_{5:0}$) along with an additional sign bit ($D_6$) to control the input polarity. Depending on the input polarity, the DAC selects the low reference ($V_{INL}=0.2\ \mathrm{V}$) for positive inputs and the high reference ($V_{INH}=0.6\ \mathrm{V}$) for negative inputs. To accommodate both positive and negative MAC results, the signal zero level is set to $\frac{V_{INH}+V_{INL}}{2}$, by applying $V_{BIAS}=0.4\ \mathrm{V}$ across the analog path of the MAC operation. The transfer function of the input DAC ($V_{DAC}$) is shown in Fig. \ref{fig:IPDAC}(b). 

\paragraph{CIM array} The CIM array consists of a 36$\times$32 ($N\times M$) grid of the proposed MDAC cells, each with a digital weight precision of B\textsubscript{W} = 6+1 bits (details in Section \ref{sec:CIM}).

\begin{figure}[t]
    \centering
    \includegraphics[width=0.9\columnwidth]{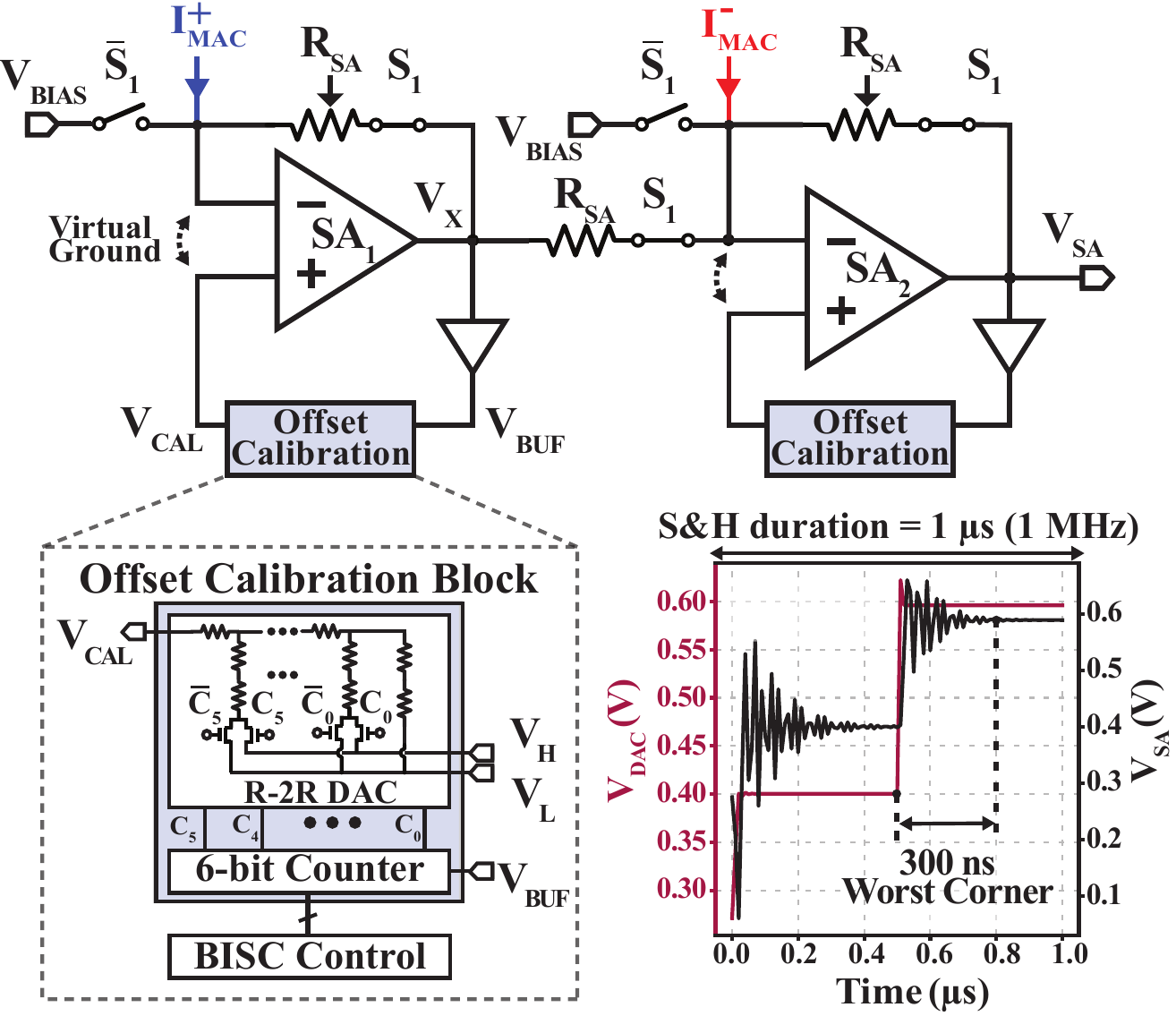}
    \caption{Schematic of the 2SA circuit, which accumulates positive and negative currents to yield a voltage output ($V_{SA}$), that fully settles with the inference period.}
    \label{fig:SA}
\end{figure}

\paragraph{Output Stage} The output stage, shown in Fig. \ref{fig:SA}, uses a two-stage Summation Amplifier (2SA) circuit to accumulate $I_{MAC}$ currents and convert them into voltage. This stage consists of two inverting summing amplifiers: SA1 sums the positive currents, while SA2 sums the negative currents and inverts the output of SA1 ($V_X$). The 2SA circuit is replicated $M$ times—one per CIM column. Each inference is computed in a single clock cycle, corresponding to the S\&H period ($T_{S\&H}=1 \mathrm{\mu s}$), during which all $M$ 2SA outputs are generated as voltages. The output voltage, $V_{SA}$, is described by:
\begin{equation}
V_{SA} = R_{SA} \cdot I_{MAC} + V_{CAL} ,
\label{VSA}
\end{equation}where $R_{SA}$ is the feedback resistor (i.e., the transresistance of the amplifier) and $V_{CAL}$ is a programmable offset (initially set to $V_{BIAS}$).

As shown in Fig. \ref{fig:SA}, the 2SA circuit fully settles within the S\&H duration. In the next clock cycle, the $M$ 2SA outputs are time-multiplexed into a flash ADC with a precision of B\textsubscript{Q} = 6 bits, which operates at, $\frac{M}{T_{S\&H}}=32\ \mathrm{MHz}$, to digitize the $V_{SA}$ signals. The theoretical MAC precision required for practical mixed-signal implementations is given by $B_D+B_W+log_2(N)$ bits \cite{ShRo22}. While reducing output precision (B\textsubscript{Q} = 6 bits) can limit maximum theoretical accuracy, it remains well-suited for DNN inference, where network robustness compensates for quantization noise. As a result, analog computation becomes more energy-efficient at reduced bit-widths, and lowering ADC resolution significantly improves power consumption, area, and speed \cite{Mur21}. The quantized output for a column ($c\in M$), denoted by $\hat{Q}$, is calculated as:

\begin{equation}
\hat{Q_c} = \frac{\,V_{SA} - V_{ADC}^{L}\,}{ \bigl(V_{ADC}^{H}-V_{ADC}^{L}\bigr) / \bigl(2^{B_{Q}} - 1\bigr)},
\label{Q1}
\end{equation}where ADC references are set to $(V_{ADC}^L,\,V_{ADC}^H)=(V_{INL},\,V_{INH})$ in default setting. 

\paragraph{BISC and system controls} BISC circuits are integrated with the 2SA stage to calibrate and correct non-linearities in the CIM core’s analog output. Since BISC occurs prior to digitizing, it ensures the correction of non-idealities prior to digitization, complementing the ADC’s role in determining final precision. The CIM core operation and BISC routines are controlled through various blocks (S\&H routine; SRAM read/write; ADC time-multiplexing; BISC control), all interfaced with the RISC-V core via AXI4-Lite. The BISC operation is detailed in Section \ref{sec:BISC}.

\section{Multi-bit Mixed-Signal CIM Core with Linear Resistors}\label{sec:CIM}

\begin{figure}[t]
    \centering
    \includegraphics[width=\columnwidth]{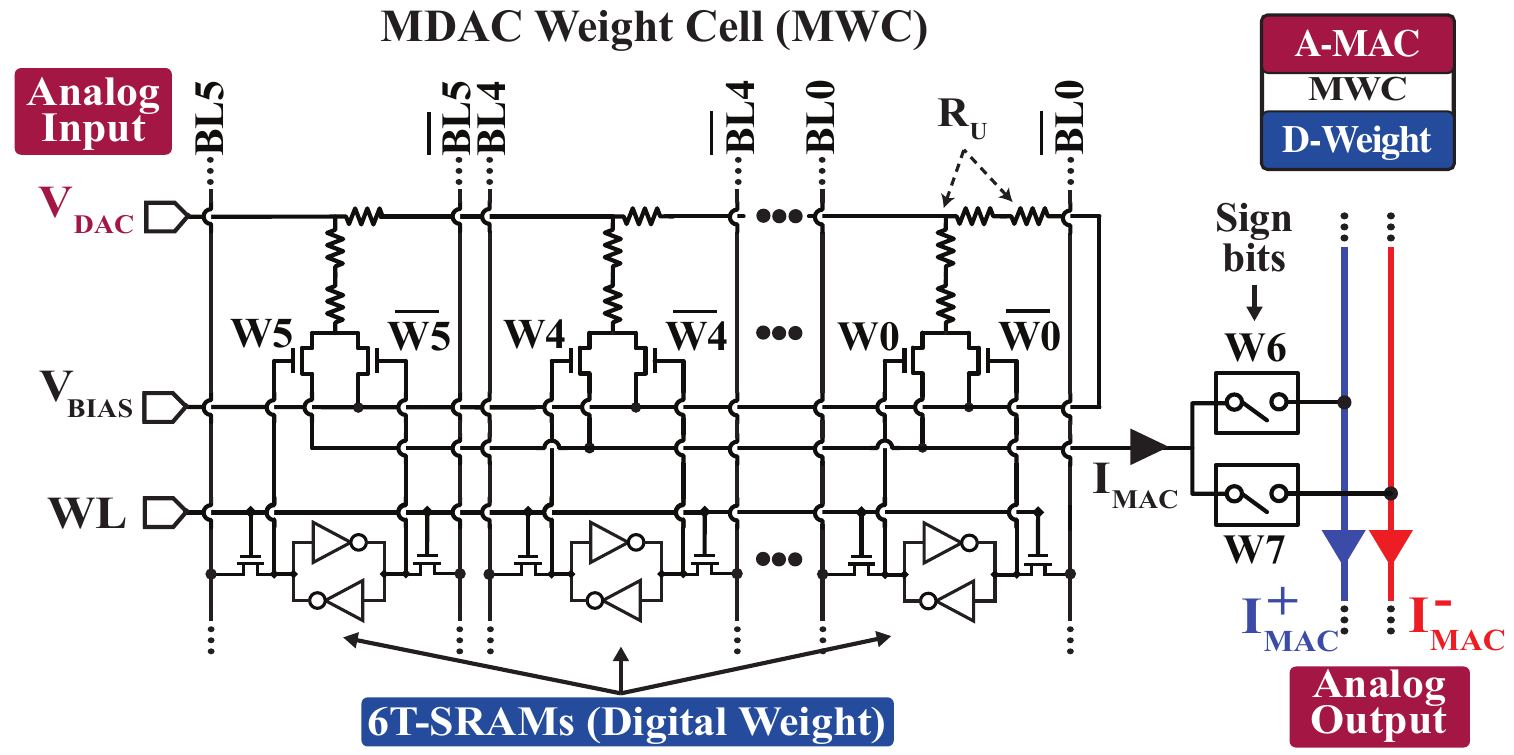}
    \caption{MDAC Weight Cell (MWC) schematic: Uses an R-2R MDAC with 6-bit weight storage and dual sign bits for polarity control.}
    \label{fig:MDAC}
\end{figure}

\noindent In this section, we detail the contribution that tackles \textbf{Ch.\circled{1}}, introducing a reliability-oriented R-2R MDAC Weight Cell (MWC) configuration, as shown in Fig. \ref{fig:MDAC}. The principle of the proposed MWC stores weight values in fast, high-density SRAM cells and employs an R-2R MDAC for efficient multi-bit multiplication. This approach reduces reliance on word-line activations and mitigates read disturbances. In addition, it improves signal stability and leverages the compact 6T-SRAM cell without compromising performance. From an integration perspective, several HDLR technologies could be used as the computing element, offering a denser implementation. For example, Mega-Ohm Resistor (MOR) \cite{SKKRB21} can integrate a 5 MΩ resistor in a 0.25 µm\textsuperscript{2} area, translating to approximately 20 MΩ per 1 µm\textsuperscript{2}, significantly higher than typical high-density polysilicon resistors. 

\subsection{MDAC Weight Cell Architecture}

The MWC architecture utilizes an R-2R MDAC topology with SRAMs for weight storage. The stored digital weight (B\textsubscript{W} = 6+1 bits) modulates the conductance of the MDAC, thereby enabling a multi-bit multiplication with the input voltage ($V_{DAC}$) to generate the output current ($I_{MAC}$). The conductance of the MWC is quantized based on $W_{5:0}$, determining the effective weight applied to the input. To handle MAC polarity, two sign bits ($W_6$ and $W_7$) are included. These allow both switches to remain off when the MWC is idle, reducing off-state leakage and minimizing offset errors at the 2SA and ADC outputs. The sign bits also control current accumulation, ensuring that like-signed currents align across MWC rows. Specifically, when $W_6$ = 1, $I_{MAC}$ is directed to the positive summation line ($I_{MAC}^{+}$), when $W_7$ = 1, it is directed to the negative summation line ($I_{MAC}^{-}$), and when both sign bits are 0, the cell remains idle. The net current, $I_{MAC}=I_{MAC}^{+}-I_{MAC}^{-}$, in a CIM column ($c\in M$) is defined as:

{\small
\begin{equation}
I_{MAC_{c}} = \sum_{i=0}^{N} \Bigl(\frac{V_{DAC}(i) - V_{BIAS}}{R_{U}}\cdot \frac{D(i)}{2^{B_{W}}}\Bigr)\cdot \bigl(W_6(i)-W_7(i)\bigr),
\end{equation}}\normalsize
where $V_{BIAS}$ is the bias voltage, set to 0.4 V to establish the zero level in the analog data path. $R_{U}$ represents the R-2R MDACs' unit resistance, $D$ is the 6-bit weight value (in integer format) corresponding to the $i^{th}$ cell, and $(W_6\,,\,W_7)$ is the state of the sign bit. 

\subsection{Extension possibility with HDLR technologies}

\begin{table}[t]
\caption{Performance estimation of the proposed CIM core with various resistive technologies}
\label{table:compR}
\centering
\resizebox{\columnwidth}{!}{%
\begin{tabular}{|c|cccc|}
\hline
 &
  \multicolumn{1}{c|}{\textbf{\begin{tabular}[c]{@{}c@{}}Polysilicon (22-nm)\\ (This work)\end{tabular}}} &
  \multicolumn{1}{c|}{\textbf{\begin{tabular}[c]{@{}c@{}}MOR \\ \cite{SKKRB21} \end{tabular}}} &
  \multicolumn{1}{c|}{\textbf{\begin{tabular}[c]{@{}c@{}}WOx \\ \cite{CKHLLH21} \end{tabular}}} &
  \textbf{\begin{tabular}[c]{@{}c@{}}RRAM (22-nm)\\ \cite{GABBBC19} \end{tabular}} \\ \hline
\textbf{\begin{tabular}[c]{@{}c@{}}Unit Resistance\\ $R_{U}$ ($M\Omega $) \end{tabular}} &
  \multicolumn{1}{c|}{0.385} &
  \multicolumn{1}{c|}{7} &
  \multicolumn{1}{c|}{28} &
  0.03 \\ \hline
\textbf{\begin{tabular}[c]{@{}c@{}}MWC Area \\ (1 bit - 6 bit)\\ ($\mu m^2$) \end{tabular}} &
  \multicolumn{1}{c|}{17 - 120} &
  \multicolumn{1}{c|}{1 - 8} &
  \multicolumn{1}{c|}{1 - 8} &
  0.05 - 0.4 \\ \hline
\textbf{\begin{tabular}[c]{@{}c@{}}Unit Current\\ ($\mu A$)$^{*}$  \end{tabular}} &
  \multicolumn{1}{c|}{2.6} &
  \multicolumn{1}{c|}{0.15} &
  \multicolumn{1}{c|}{0.036} &
  33 \\ \hline
\begin{tabular}[c]{@{}c@{}} \textbf{Area Improv.}\\ \textbf{Power Improv.}\end{tabular} &
\multicolumn{1}{c|}{\begin{tabular}[c]{@{}c@{}} \textbf{Baseline} \end{tabular}} &
\multicolumn{1}{c|}{\begin{tabular}[c]{@{}c@{}} 14$\times$\\17$\times$ \end{tabular}} &
\multicolumn{1}{c|}{\begin{tabular}[c]{@{}c@{}} 14$\times$\\70$\times$ \end{tabular}} &
\multicolumn{1}{c|}{\begin{tabular}[c]{@{}c@{}} 225$\times$\\0.08$\times$ \end{tabular}} \\ \hline

\end{tabular}%
}
\RaggedRight
\begin{threeparttable}
   \begin{tablenotes}
      \footnotesize
      \item  $^{*}$ per MWC, assuming 1V operation.
   \end{tablenotes}
\end{threeparttable}
\end{table}

Although the proof-of-concept SoC demonstrates the functionality of the proposed design concepts, the use of polysilicon resistors impacts the area and energy efficiency. However, alternative resistive technologies, such as HDLR materials, offer the potential to significantly improve these metrics. To explore this potential, we evaluated the performance of the proposed MWC using different resistive technologies, taking the unit resistance ($R_U$) value of the polysilicon-based design as a baseline. For HDLR technologies, we assume that the R-2R resistor network can be post-processed atop an SRAM cell, with a resistor value of 3$R_U$ occupying roughly 1 µm\textsuperscript{2}. This approach could enable a 128$\times$128 MWC cell array to fit within the same 0.14 mm\textsuperscript{2} footprint used by the proof-of-concept SoC. Table \ref{table:compR} summarizes these comparisons and their implications. 

\paragraph{Polysilicon Resistors}
The baseline design uses high-density polysilicon resistors fabricated in a 22-nm FD-SOI process. This implementation supports a 36$\times$32 MWC array and provides a reference for evaluating other technologies.

\paragraph{Mega-Ohm Resistor (MOR) Technology}
MOR technology \cite{SKKRB21} can achieve a resistance of 5 MΩ in a 0.25 µm\textsuperscript{2} area, translating to approximately 20 MΩ per 1 µm\textsuperscript{2}. For this study, we assume an $R_U$ of 7 MΩ, reducing current draw to 150 nA per MWC. This results in a 14$\times$ improvement in both throughput and area efficiency and a 17$\times$ reduction in power consumption, excluding peripherals.

\paragraph{WOx Resistor Technology}
WOx-based resistors \cite{CKHLLH21} offer an even higher resistance density. With an estimated $R_U$ of 28 MΩ, each MWC draws only 36 nA of current, yielding up to 70$\times$ energy efficiency improvement compared to the baseline, excluding peripheral components.

\paragraph{RRAM Integration}
We also evaluated RRAM, fabricated using the same 22-nm node as the proof-of-concept design \cite{GABBBC19}. While RRAM provides dense non-volatile memory integration, its current draw per cell (33 µA) limits its energy efficiency, resulting in significantly higher power consumption compared to other technologies.

\section{Open-source system simulation and measurement framework} \label{sec:git}

\begin{figure}[t]
    \centering
    \includegraphics[width=\columnwidth]{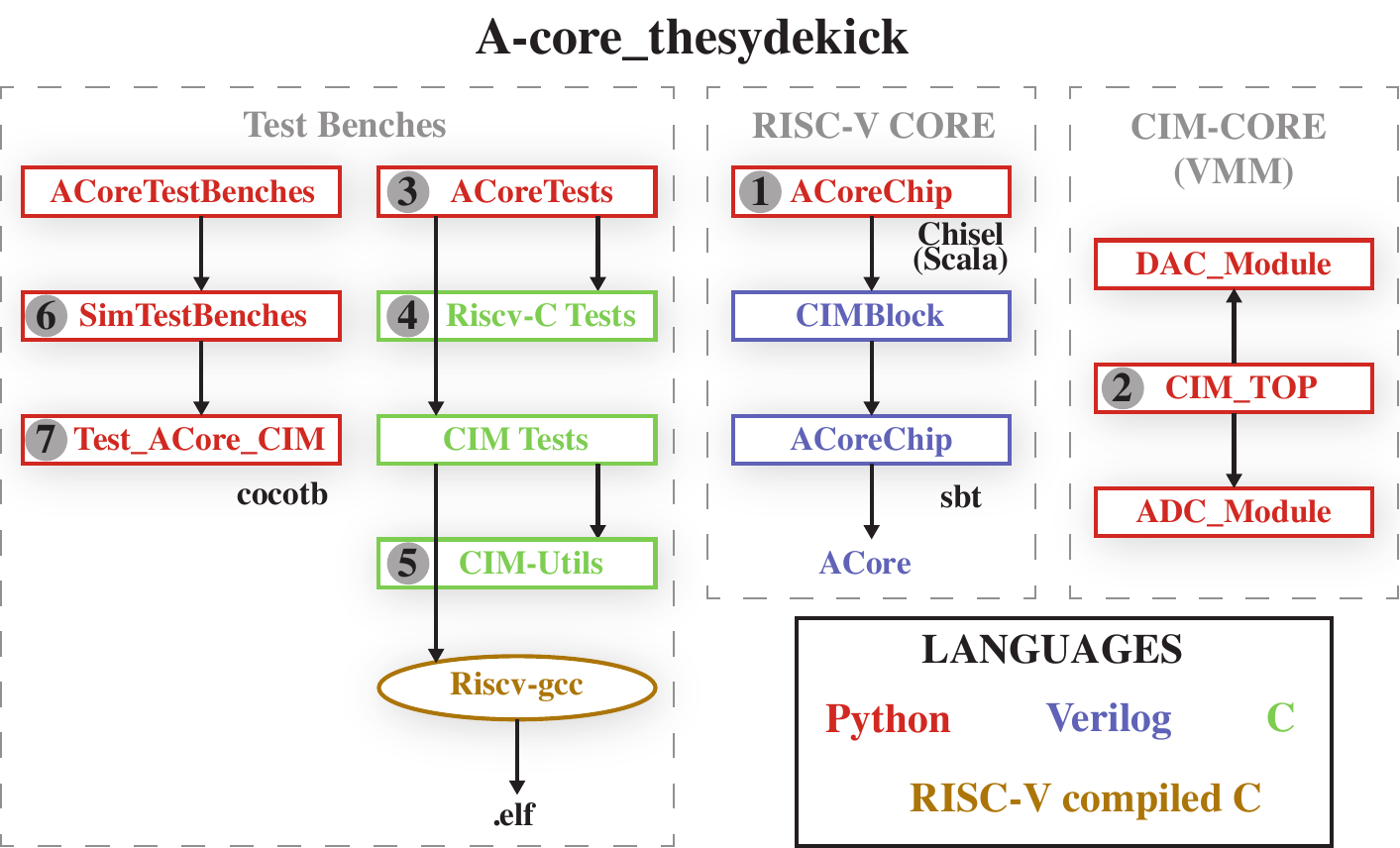}
    \caption{Illustration of the open-source simulation framework of the proposed Acore-CIM SoC, including the A-core RISC-V core (implemented in RTL), the CIM core (modeled in Python), and various test benches.}
    \label{fig:Model}
\end{figure}

\noindent In this section, we detail the contribution that tackles \textbf{Ch.\circled{2}}. To promote CIM-based systems, extend SoC testing, and enable future integration, we developed a complete open-source simulation framework for the proposed SoC\footnote{The complete source-code is available at: https://gitlab.com/a-core/}. It relies on the System Development Kit (SyDeKick) environment for system modeling\footnote{https://github.com/TheSystemDevelopmentKit}. This simulation framework supports the implementation of simulation, measurement, and test routines for complex SoCs, where each block can be modeled at different abstraction levels (behavioral, RTL, schematic, or post-layout), tested, and subsequently replaced with the physical design while retaining the same measurement setup. It also supports efficient co-simulation of software and hardware to prepare programming or test routines. All measurements detailed in the following subsections use this framework.

\subsection{Framework description} 

Fig. \ref{fig:Model} depicts the block diagram of the proposed simulation framework, consisting of three main components: the A-core RISC-V core in RTL, the CIM core in Python, and various test benches. The processor and accelerator communicate over the same AXI4-Lite bus as the fabricated SoC. This interface is implemented in the "$CIM_{Block}$" file, along with the processor’s Verilog code. Since the CIM core is modeled in Python, a co-simulation between the RTL (processor and AXI4-Lite interface) and Python (CIM) is necessary, realized with CoCoTB\footnote{https://www.cocotb.org/}. For this purpose, the “ACoreTestBenches” file sets up the framework to halt the RTL simulation at each clock cycle and run the “$CIM_{top}$” file for CIM simulation, allowing bidirectional data transfer via “$CIM_{Block}$”. This approach avoids long CIM simulation times while using the same test bench (compiled from C code by gcc) as the physical chip, enabling efficient testing of various programs on the system.

\subsection{Detailed simulation operation}

Fig. \ref{fig:Model} details all the main files used in the simulation framework. The following explanation can be linked with their respective markers, denoted with a circle \circled{X}. The AcoreChip file includes the complete RTL code for the processor in Chisel and the AXI4 interface to the CIM core \circled{1}. This code is compiled into Verilog via sbt, a Scala-based build tool. The CIM core is modeled in Python \circled{2}, with separate models for data converters to facilitate independent testing and rapid prototyping. The model interfaces with the “$CIM_{Block}$” via the AXI4 interface, i.e., data can be sent to and from the Python model during co-simulation. Using SydDeKick functionalities, all CIM core blocks can be replaced by their hardware entities, either at the simulation level (netlist or RTL code) or as the physical SoC itself. The "ACoreTests" file contains all system tests \circled{3}, including standard RISC-V tests for the processor \circled{4} and tests for CIM core utility functions \circled{5}. Each test's C-code can be written and compiled with the RISCV-gcc compiler, producing an \textit{.elf} instructions file, which can also be loaded into the SoC chip's program memory. The "ACoreTestBenches" file compiles all simulation source files and initiates the simulation with CoCoTB \circled{6}, generating the co-simulation with the RTL test bench and the compiled RISCV-gcc file running the processor code \circled{7}. The same test benches are used both in simulation and can be directly ported to the physical design, ensuring consistency between pre-silicon and post-silicon validation.

\section{RISC-V Controlled Built-In Self-Calibration}\label{sec:BISC}

\noindent In this section, we tackle \textbf{Ch.\circled{3}}. As introduced earlier, mixed-signal CIM cores suffer from errors—stemming from process variations, device mismatches, parasitics, and amplifier non-idealities—that degrade the accuracy of MAC operations. We propose an on-chip BISC to systematically identify and calibrate these errors. BISC methodology is fully controlled by the RISC-V core and can be automated. This enables real-time error correction without significant system overhead. The complete BISC implementation is available in our open source framework.

\subsection{Error Modeling and Correction Strategy}

Without errors, the nominal output of the SA is given by $V_{SA,nom} = R_{SA}\,I_{MAC} + V_{CAL}$. In reality, the SA is affected by analog–domain non-idealities that introduce both gain and offset errors. To model them, let us consider a gain error factor $\alpha_{A}$ (ideally 1) and an additive offset error $\beta_{A}$, which modify the output as:
\begin{equation}
    V_{SA,act} = \alpha_{A}\, R_{SA}\, I_{MAC} + V_{CAL} + \beta_{A}.
    \label{VSA_act}
\end{equation}

To determine the appropriate calibration values, we introduce gain and offset correction factors, resp. $\gamma_{SA}$ and $\Delta_{CAL}$. First, to compensate the gain error, we define a correction factor $\gamma_{SA}$ such that $\gamma_{SA}\, \alpha_{A} = 1$, yielding a calibrated transresistance: 
    \begin{equation}
        R_{SA}' = \gamma_{SA}\, R_{SA} = \frac{R_{SA}}{\alpha_{A}}.
        \label{RS_new_correction}
    \end{equation}

Second, we cancel the offset error by setting $\Delta_{CAL} + \beta_{A} = 0$, which leads to a corrected calibration voltage:
    \begin{equation}
        V_{CAL}' = V_{CAL} + \Delta_{CAL} = V_{CAL} - \beta_{A}.
        \label{VCAL_new_correction}
    \end{equation}

Hence, the application of the corrected values $R_{SA}'$ and $V_{CAL}'$ to the actual SA output \eqref{VSA_act} via the BISC circuits yields a calibrated SA output $V_{SA,BISC}$ that restores the ideal behavior of $V_{SA,nom}$ over the full range of $I_{MAC}$.

To correct these errors, Fig.~\ref{fig:SA} shows that our 2SA circuit incorporates a digital potentiometer in its negative feedback path and a 6-bit voltage-mode R-2R DAC (controlled by an up-counter) in the positive feedback loop. This design renders $R_{SA}$ and $V_{CAL}$ both tunable, as detailed in the next subsection.

\subsection{ADC Integration and Combined Error Analysis}

When the CIM core is integrated into a system, the SA outputs are quantized by an ADC. Hence, in a full system analysis, the ADC's error needs to be considered. The nominal ADC transfer function is defined as:
\begin{align}
Q_{\text{nom}} &= C_{ADC}\,\Bigl[ R_{SA}\,I_{MAC} + V_{CAL} - V_{ADC}^{L} \Bigr],
\label{eq:Q_nom}
\end{align}
where $C_{ADC}$ is the constant ADC conversion factor defined as $\bigl(2^{B_Q}-1\bigr)/\bigl(V_{ADC}^{H}-V_{ADC}^{L}\bigr)$. In addition to the SA non-idealities, the ADC itself exhibits errors, which we also model with digital gain $\alpha_{D}$ and offset $\beta_{D}$ errors. As a result, the system's response is expressed in a linear form as: 
\begin{align}
\hat{Q}_{\text{act}} &= \alpha_{D}\, C_{ADC}\,\Biggl[ \alpha_{A}\,R_{SA}\,I_{MAC} \nonumber\\[1mm]
&\quad +\, V_{CAL} + \beta_{A} + \frac{\beta_{D}}{\alpha_{D}\,C_{ADC}} - V_{ADC}^{L} \Biggr].
\label{eq:Q_act}
\end{align}

Assuming that we can only measure the system's response after the ADC, the system's linear error can be effectively expressed by modeling the measured ADC output as:
\begin{align}
\hat{Q}_{\text{act}} = \hat{g}_{\text{tot}}\,Q_{\text{nom}} + \hat{\epsilon}_{\text{tot}}\, ,
\label{eq:Qact_eql_Qnom}
\end{align}where $\hat{g}_{\text{tot}}$ and $\hat{\epsilon}_{\text{tot}}$ denote resp. the combined gain and offset errors, arising from the complete CIM column, including the ADC. By comparing the coefficients of $I_{MAC}$ and the constant terms in \eqref{eq:Q_nom} and \eqref{eq:Q_act}, and by setting $V_{\text{CAL}} = V_{\text{ADC}}^L$, we obtain
\begin{align}
    \hat{g}_{\text{tot}} &= \alpha_{A}\,\alpha_{D}, \quad \text{and}\quad 
    \hat{\epsilon}_{\text{tot}} = \alpha_{D}\,C_{ADC}\,\beta_{A} + \beta_{D}.
    \label{eq:gtot_eps_final}
\end{align}
The Equations in \eqref{eq:gtot_eps_final} indicate that the overall gain error is the product of the analog and digital gain errors, while the total offset error is the sum of the digital offset and the analog offset scaled by both the ADC conversion factor and the digital gain.

Assuming that the ADC has been characterized independently (i.e., its gain error $\alpha_{D}$ and offset error $\beta_{D}$ are known), the analog errors ($\alpha_{A}$ and $\beta_{A}$) of the CIM core outputs can be extracted from the Equations in \eqref{eq:gtot_eps_final} as: 
\begin{align}
    \alpha_{A} &= \frac{\hat{g}_{\text{tot}}}{\alpha_{D}}\ ,\quad \text{and}\quad
    \beta_{A} = \frac{\hat{\epsilon}_{\text{tot}} - \beta_{D}}{\alpha_{D}\,C_{ADC}}\ .
    \label{eq:alphaA_betaA_ext}
\end{align}
These values serve as correction factors to compute the final calibrated parameters and used to update the tunable elements in the 2SA circuit. Specifically, using the corrections from \eqref{RS_new_correction} and \eqref{VCAL_new_correction}, we obtain: 
\begin{align}
    R_{\text{SA}}' &= \frac{\alpha_{D}\,R_{\text{SA}}}{\hat{g}_{\text{tot}}}\ ,\quad \text{and}\quad
    V_{\text{CAL}}' = V_{\text{CAL}} - \frac{\hat{\epsilon}_{\text{tot}} - \beta_{D}}{\alpha_{D}\,C_{ADC}}\ .
    \label{eq:final_rsa_vcal}
\end{align}

\subsection{BISC routine}
Once we determined the theoretical correction factors, the BISC routine must apply the correct test vectors to extract the actual errors (the "\textit{online characterization}" phase) and then calculate the optimal correction factors according to the previous analysis (the "\textit{online correction}" phase).

\subsubsection{Online characterization Phase}
To characterize the complete CIM column (the combined 2SA+ADC chain), we apply a set of known MAC operations, leading to a known accumulated current $\{I_{MAC}\}$. We then measure the corresponding digital outputs $\hat{Q}_{\text{act}}$, which are then compared with the nominal outputs $\{Q_{\text{nom}}\}$. Using the linear relationship of  \eqref{eq:Qact_eql_Qnom}, we estimate the measured gain and offsets with a least-squares fit over $Z$ test points, yielding:
\begin{align}
    \hat{g}_{\text{tot}} &= \frac{Z\,\sum \left(Q_{\text{nom}}\,\hat{Q}_{\text{act}}\right) - \sum Q_{\text{nom}}\,\sum \hat{Q}_{\text{act}}}
    {Z\,\sum \left(Q_{\text{nom}}\right)^2 - \left(\sum Q_{\text{nom}}\right)^2}\ ,
    \label{eq:LS_g}\\[1mm]
    \hat{\epsilon}_{\text{tot}} &= \frac{\sum \hat{Q}_{\text{act}} - \hat{g}_{\text{tot}}\sum Q_{\text{nom}}}{Z}\ .
    \label{eq:LS_eps}
\end{align}

Selecting the test set size ($Z$) involves a critical trade-off. Indeed, a full-range sweep, measuring every LSB, yields high accuracy and reveals non-linearity, but is too time-consuming for large arrays. Conversely, a minimal two-point method allows rapid calibration with low overhead yet risks missing mid-range distortions and is more noise-prone. To strike a balance between these extremes, we use a small set of 4–8 equally spaced test vectors across the dynamic range and perform a least-squares regression. Additionally, the BISC allows measuring the output multiple times at each test point to average out random errors, such as thermal and flicker noise.

\subsubsection{Online correction Phase}
Based on the estimated $\hat{g}_{\text{tot}}$ and $\hat{\epsilon}_{\text{tot}}$, the SA trimming values are updated as per Equations \eqref{eq:final_rsa_vcal}. Specifically, the digital potentiometer in the negative feedback path is fine-tuned for gain correction, while the calibration DAC output in the positive feedback loop is adjusted for offset correction. These updates compensate for systematic, linear errors in the analog signal path.

\subsection{Additional Considerations for BISC}

\paragraph{ADC Clipping} Residual errors may cause $V_{SA,\text{act}}$ to exceed the ADC limits, leading to saturation. To ensure valid calibration data and prevent clipping, the ADC reference voltages $V_{ADC}^{L}$ and $V_{ADC}^{H}$ can be adjusted so that the full dynamic range—including error margins—remains within the ADC’s linear region. For example, if the nominal SA output $V_{SA,nom}$ is in the range $0.4\!-\!0.6\,\mathrm{V}$ and a $\pm5\%$ margin is expected, the ADC references are set to approximately $(V_{ADC}^L,\,V_{ADC}^H) \approx (0.39\,\mathrm{V},\,0.61\,\mathrm{V})$, ensuring that all test points remain in-range.

\begin{figure}[t]
    \centering
    \includegraphics[width=\columnwidth]{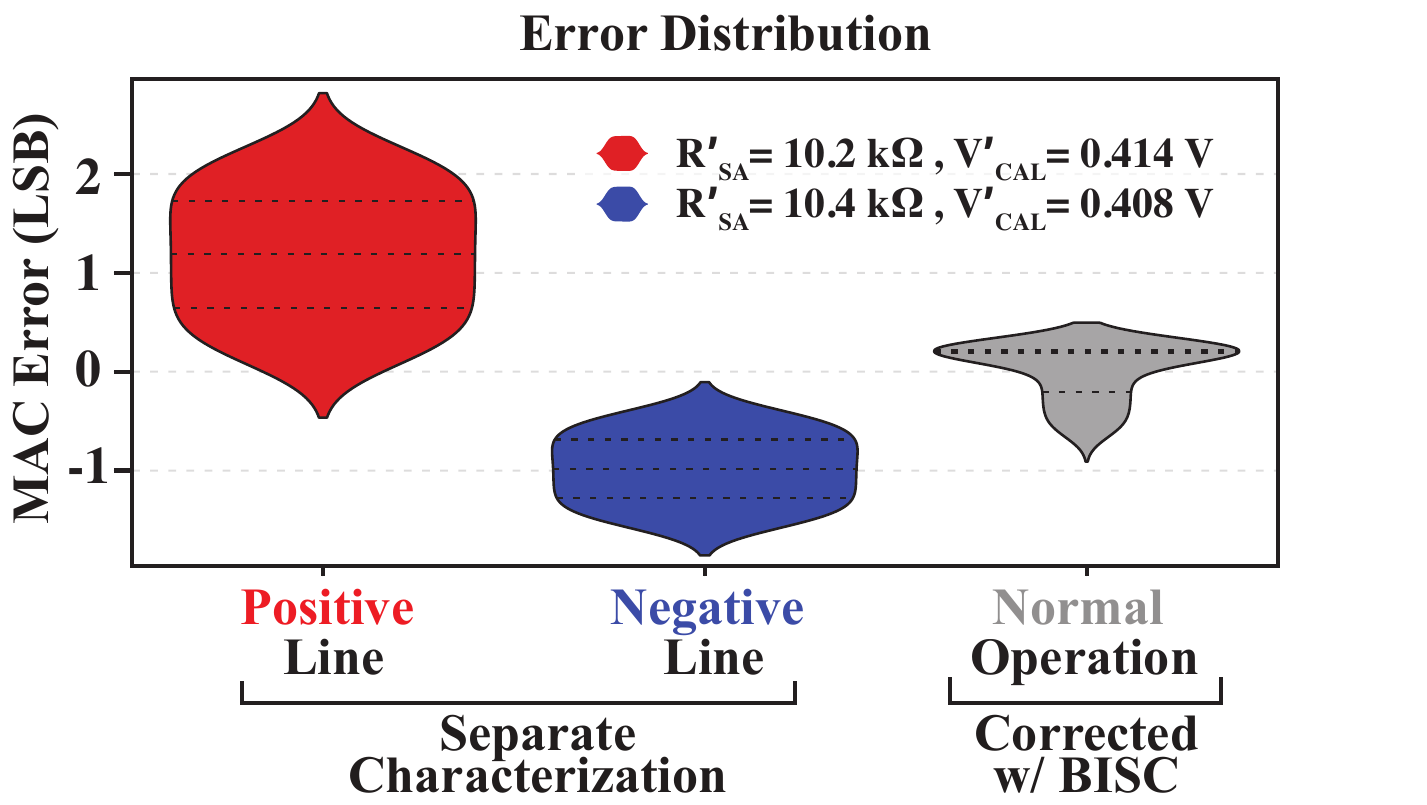}
    \caption{Error distributions for a selected CIM column during characterization phase and after BISC calibration. Default settings w/o BISC: $R_{SA}$ = 10.7 kΩ, and $V_{CAL}$ = 0.4 V}
    \label{fig:BISC_Calib}
\end{figure}

\paragraph{Separate calibration} 
In the proposed CIM core, SA1 and SA2 may exhibit distinct non-linearities due to device mismatch, process variations, and parasitic elements. Thus, we independently measure and correct offset and gain errors in SA1 and SA2, ensuring precise compensation for each line’s unique non-idealities and asymmetries. Fig. \ref{fig:BISC_Calib} shows this for a selected CIM column. The “Positive Line” and “Negative Line” distributions reveal distinct amplifier error profiles before calibration. The BISC routine then determines separate trimming values for SA1 and SA2. The “Normal Operation” distribution demonstrates how these adjustments significantly reduce errors, underscoring the need for calibration to ensure precise, reliable mixed-signal CIM performance.

The pseudocode of the BISC routine is listed in Algorithm \ref{alg:BISC}. The BISC routine is flexible and can be executed after a system reset, following a classification task, or periodically at predefined intervals, depending on application needs. This demonstrates the seamless integration of BISC with the RISC-V core, ensuring efficient real-time calibration if needed.

\begin{algorithm}[htbp]
\caption{BISC Calibration Routine}
\label{alg:BISC}
\textbf{Initialization:} \\
\textbf{Set} \texttt{RISC-V controls}. \\
\textbf{Initialize Calibration Variables:} \\
\Indp
- $V_{CAL} \gets \frac{V_L + V_H}{2}$, \& $R_{SA} \gets \frac{R_U}{N}$. \\
- $V_{ADC}^{L} \gets 0.95\,V_{ADC}^{L,\text{default}}$. \\
- $V_{ADC}^{H} \gets 1.05\,V_{ADC}^{H,\text{default}}$. \\
\Indm
\textbf{Store ADC Parameters:} $(\alpha_D,\,\beta_D,\,C_{ADC})$. \\
\textbf{Define Test Vectors:} $\{V_{t}[Z],W_t[Z]\}$ and $Q_{exp}$. \\

\BlankLine
\textbf{Characterization Phase:} \\
\ForEach{column $c$ in CIM array}{
    - Write $W_t \gets W_{\max}.\ \ V_{t} \gets$ stepped input.\\
    - $\{V_{t}[Z],W_t[Z]\} \to  \{I_{MAC}[Z]\}$.\\
    - Measure $\hat{Q}_{act}$ per column. \\
    - Compute: $\hat{g}_{tot} \gets$ per Eq.\eqref{eq:LS_g}, $\hat{\epsilon}_{tot} \gets$ per Eq.\eqref{eq:LS_eps}.
}
%\BlankLine
\textbf{Correction Phase:} \\
\ForEach{column $c$ in CIM array}{
    \textbf{Gain Correction:} 
    $R_{SA}' \gets \dfrac{\alpha_D\,R_{SA}}{\hat{g}_{tot}}$.\\
    %\BlankLine
    \textbf{Offset Correction:} 
    $V_{CAL}' \gets V_{CAL} - \dfrac{\hat{\epsilon}_{tot}-\beta_D}{\alpha_D\,C_{ADC}}$.\\
    %\BlankLine
}
Update BISC circuits with $R_{SA}' \& V_{CAL}'$.
\end{algorithm}

\section{Measurement Results of The A-core CIM Prototype}\label{sec:measurements}

\begin{table*}[t]
    \centering
    \caption{Comparison with state-of-the-art CIM accelerators including calibration features}
    \label{tab:comp}
    \renewcommand{\arraystretch}{1.3}
    \resizebox{\textwidth}{!}{
    \begin{tabular}{c c c c c c c c c c c}
        \toprule[1.5pt]
\textbf{} &
  JSSC'24 \cite{PLWLCL24} &
  JSSC'21 \cite{SENAS21} &
  JSSC'24 \cite{SSC24} &
  VLSI'22 \cite{CJSLZ22} &
  JSSC'23 \cite{HWHHS22} &
  \textbf{This SoC} \\
  \midrule[1pt]
  
\textbf{Technology (nm)} & 180 @ 1.8\,V & 7 @ 0.8\,V & 28 @ 1\,V & 65 @ 1.2\,V & 22 @ 0.8\,V & 22 @ 0.8\,V \\
  \midrule[1pt]

\textbf{\begin{tabular}[c]{@{}c@{}}CIM Technique\\ (Weight Memory) \end{tabular}} &
\begin{tabular}[c]{@{}c@{}}Current-based DAC \\ (SRAM)\end{tabular} &
8T-SRAM &
\begin{tabular}[c]{@{}c@{}}Charge Accum. MAC \\ (SRAM)\end{tabular} &
\begin{tabular}[c]{@{}c@{}}1T1R \\ (RRAM)\end{tabular} &
\begin{tabular}[c]{@{}c@{}}1T1R SLC\\ (RRAM)\end{tabular} &
\begin{tabular}[c]{@{}c@{}}R-2R MDAC \\ (SRAM)\end{tabular} \\ 
\midrule[1pt]

\textbf{Calibration} & 
Process Variation & 
Transistor Non-linearity & 
PVT Variation &  
ReRAM Non-linearity &  
Timing Variations &  
CIM Non-idealities \\ 
\midrule[1pt]
\begin{tabular}[c]{@{}c@{}}\textbf{Calibration Features} \\ \textbf{(On/off-chip)}\end{tabular} & 
\begin{tabular}[c]{@{}c@{}}Weight Calibration \\ External Characterization \\ Hybrid\end{tabular} & 
\begin{tabular}[c]{@{}c@{}}Non-linearity Modeling \\ Re-training \\ Off-chip\end{tabular} & 
\begin{tabular}[c]{@{}c@{}}MAC Bias Adjustment \\ Real-time \\ On-chip\end{tabular} &  
\begin{tabular}[c]{@{}c@{}}Write-Verify Algorithm \\ Adaptive Programming \\ On-chip\end{tabular} &  
\begin{tabular}[c]{@{}c@{}}Timing Table \\ Pre-charge Calibration \\ On-chip\end{tabular} &  
\begin{tabular}[c]{@{}c@{}}Offset/Gain Compensation \\ Accumulation Line Tuning \\ On-chip\end{tabular} \\ 
\midrule[1pt]

\textbf{\begin{tabular}[c]{@{}c@{}}CIM Inference freq. (MHz)\\ Precision (I/P:W:O/P)  \end{tabular}} &
  \begin{tabular}[c]{@{}c@{}}0.83 \\ 4:4:Analog\end{tabular} &
  \begin{tabular}[c]{@{}c@{}}182\\ 4:4:4 \end{tabular} &
  \begin{tabular}[c]{@{}c@{}}200\\ 4:5:Analog \end{tabular} &
  \begin{tabular}[c]{@{}c@{}}63\\ 8:4:8 \end{tabular} &
  \begin{tabular}[c]{@{}c@{}}70\\ 8:8:19(IVTC) \end{tabular} &
  \begin{tabular}[c]{@{}c@{}}1 \\ 7:7:6 \end{tabular} \\ 
\midrule[1pt]

\textbf{Norm. Throughput (1b-GOPS)\textsuperscript{*}} & 6.8 & 1489 & 2520/path & 5741/tile & 9102 & 113 \\
\midrule[1pt]

\textbf{Norm. Energy Efficiency (1b-TOPS/W)} & 107.5 & 1.05 & 1105/path & 2035/tile & 0.64 & 6.65 \\
\midrule[1pt]

\textbf{Norm. Area Efficiency (1b-TOPS/mm\textsuperscript{2})} & 0.0525 & 465.3 & 10.6/path & 18.3/tile & 0.5057 & 0.155 \\
  \midrule[1pt]

\textbf{\begin{tabular}[c]{@{}c@{}}Inference Accuracy (\%)\\ Neural Network Model\end{tabular}} &
  \begin{tabular}[c]{@{}c@{}}95.69\\ MNIST on MLP\end{tabular} &
  \begin{tabular}[c]{@{}c@{}}96.5\\ MNIST on MLP\end{tabular} &
  \begin{tabular}[c]{@{}c@{}}98.1\\ MNIST on CNN\end{tabular} &
  \begin{tabular}[c]{@{}c@{}}96.8\\ MNIST on LeNet-1\end{tabular} &
  \begin{tabular}[c]{@{}c@{}}91.74\\ CIFAR-10 on ReNet-20\end{tabular} & 
  \begin{tabular}[c]{@{}c@{}}92.33\\ MNIST on MLP\end{tabular} \\
    \bottomrule[1.5pt]
    \end{tabular}
    }

\RaggedRight
\begin{threeparttable}
   \begin{tablenotes}
      \footnotesize
      \item  \textsuperscript{*} The normalized throughput (1b-GOPS) $= \eta_{MAC}\times (B_{D}\times B_{W})_{inf}\times f_{inf}$

   \end{tablenotes}
\end{threeparttable}

\end{table*}

\noindent Fig. \ref{fig:CHIP_BD}(b) shows a microphotograph of the prototype SoC chip, designed and fabricated using 22-nm FD-SOI technology. The annotated image highlights key chip sections and their corresponding locations in the block diagram. The CIM core occupies 0.73 mm\textsuperscript{2} and integrates a 36×32 MWC array along with a 2SA stage, BISC, 36×1 input DACs, SRAM read/write controls, codecs, and a time-multiplexed Flash ADC. The RISC-V core and other digital circuits together cover an additional 1.14 mm\textsuperscript{2}. The power distribution of the prototype
chip is illustrated in Fig. \ref{fig:CHIP_BD}(c). The measurement setup involves a custom PCB connected to a ZC706 FPGA for control, with power supplied and measured using Keysight N6705C DC Power Analyzers.

\begin{figure*}[t]
    \centering
    \includegraphics[width=\textwidth]{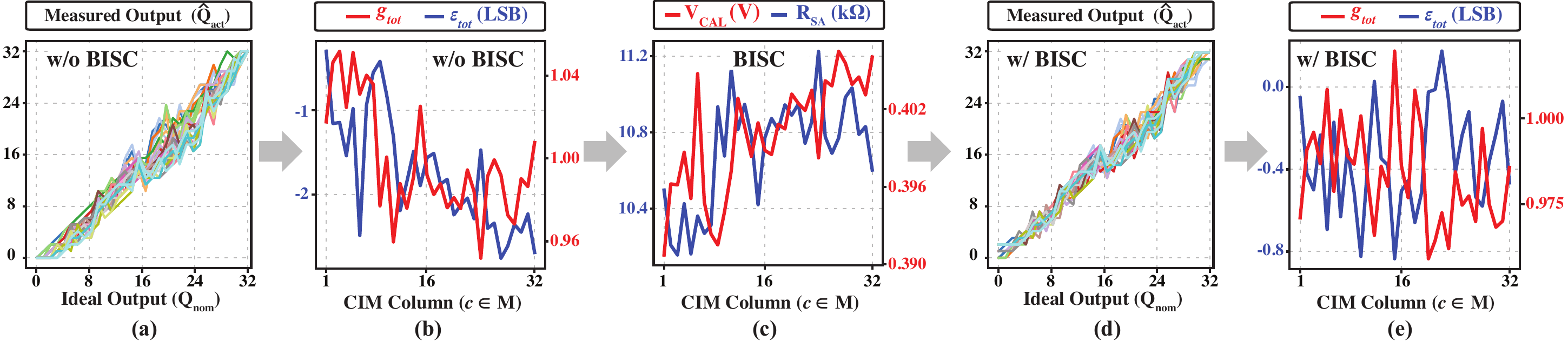}
    \caption{(a) Uncalibrated MAC outputs across CIM columns. (b) Extracted per-column gain ($g$) and offset ($\epsilon$) errors. (c) BISC-calibrated $R_{SA}$ and $V_{CAL}$ values. (d) Calibrated MAC outputs. (e) Residual gain and offset errors after calibration.}
    \label{fig:meas_bisc}
\end{figure*}

\subsection{CIM variation and calibration}

Figure \ref{fig:meas_bisc} demonstrates the progression from uncalibrated to calibrated MAC outputs. Fig. \ref{fig:meas_bisc} (a) shows the default (uncalibrated) MAC outputs with significant column-wise variations, arising from all variations experienced by the CIM core. Fig. \ref{fig:meas_bisc} (b) presents the corresponding per-column gain ($g_{tot}$) and offset ($\epsilon_{tot}$) errors extracted from these measurements. In Fig. \ref{fig:meas_bisc} (c), the BISC-determined corrections for $R_{SA}$ and $V_{CAL}$ are displayed. With these corrections applied, Fig. \ref{fig:meas_bisc} (d) shows improved MAC outputs with reduced variations, while Fig. \ref{fig:meas_bisc} (e) confirms that the residual gain and offset errors are considerably reduced.

\begin{figure}[t]
    \centering
    \includegraphics[width=\columnwidth]{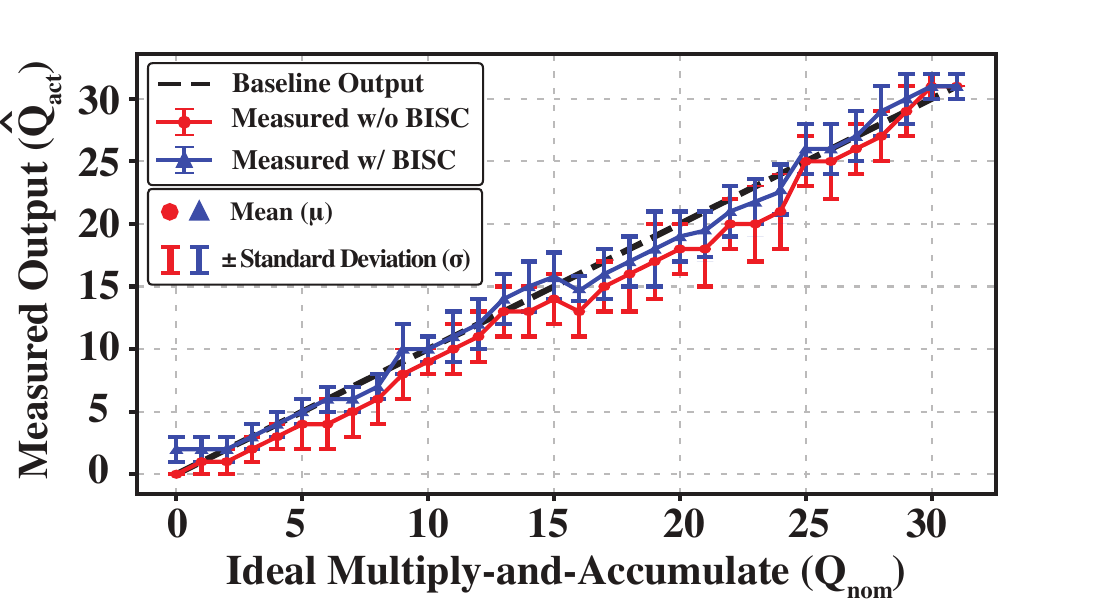}
    %\vspace{-5pt}
    \caption{A comparison of spatial variation enhancement across CIM columns without and with BISC.}
    %\vspace{-7pt}
    \label{fig:Variation}
\end{figure}

Fig. \ref{fig:Variation} compares the average CIM macro outputs with the ideal MAC values, for both uncalibrated and BISC-calibrated settings. The uncalibrated results exhibit a notable offset from the ideal curve, whereas the BISC-calibrated outputs align much more closely, demonstrating reduced spatial variation and enhanced accuracy.

\subsection{Compute SNR Evaluation}

The compute SNR of the CIM core is evaluated following the method in \cite{Shan22}, where the per-column SNR, $\mathrm{SNR}_{c}$, is defined by:
\begin{equation}
    \mathrm{SNR}_{c} 
    = \frac{\sigma_{Q_{\mathrm{nom}}}^2}{\sigma_{e}^2} ,
    \quad\text{with}\quad
    e = Q_{\mathrm{nom}} - \hat{Q}_{\mathrm{act}}.
\label{eq:SNR_compute}
\end{equation} Here, $\sigma_{Q_{\mathrm{nom}}}^2$ represents the variance of the ideal MAC output and $\sigma_e^2$ the variance of the error signal, where $e$ is the difference between the expected and measured outputs.

\begin{figure}[t]
    \centering
    \includegraphics[width=\columnwidth]{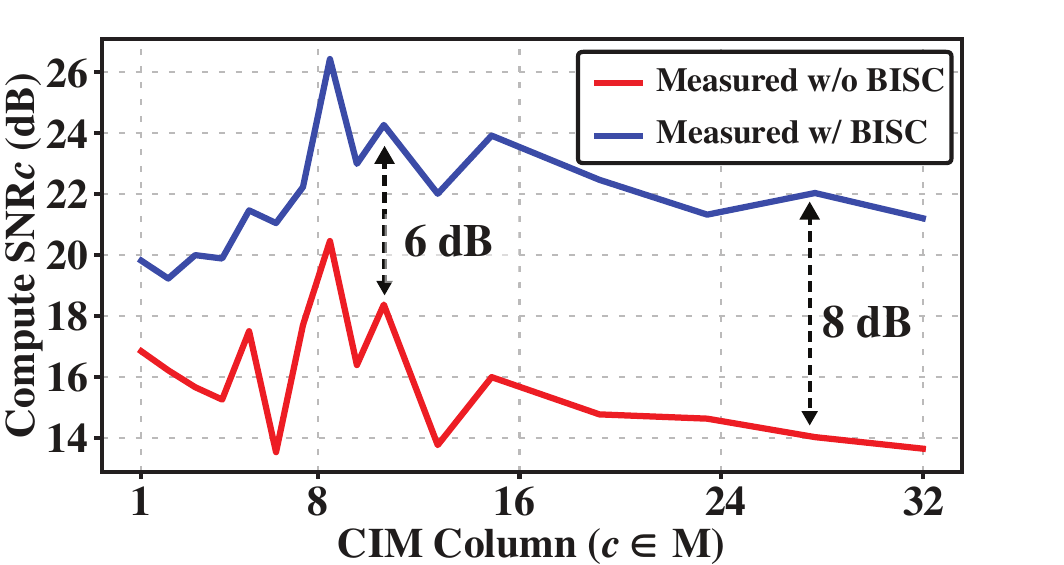}
    %\vspace{-5pt}
    \caption{Compute SNR boost across CIM columns with BISC, achieving an average boost of 6 dB.}
    %\vspace{-7pt}
    \label{fig:SNR}
\end{figure}

Fig. \ref{fig:SNR} shows the measured SNR across the CIM columns, comparing uncalibrated operation and BISC-calibrated results. As seen, BISC
allows up to 8\,dB SNR improvement on the computational SNR (6 dB on average, with improvements for every column). Hence, the system’s ENOB can attain 4 bits (with average increasing from 2.3 to 3.3 bits).

\subsection{DNN Demonstration}
The Acore-CIM chip is evaluated using an MLP model (784I-72H-10O) for MNIST digit classification \cite{deng2012mnist}. In simulation, the network achieves a baseline accuracy of 94.23\%. However, on-chip—where the CIM core executes the dot-product MAC operations and the RISC-V core accumulates intermediate results and applies bias and activation—the impact of non-idealities reduces the accuracy to 88.7\%. After characterizing and correcting offset and gain errors via BISC (i.e., adjusting $R_{SA}$ and $V_{CAL}$), the accuracy improves to 92.33\%, emphasizing that the calibration routine significantly narrows the accuracy gap between simulation and silicon. At full utilization, the CIM core consumes 16.9 nJ of energy per inference cycle.

\subsection{Chip performance summary}
Table \ref{tab:comp} compares the performances of our SoC with state-of-the-art SRAM- and RRAM-based CIM designs. The normalized CIM's throughput is defined as $1b-GOPS = \eta_{MAC}\times (B_{D}\times B_{W})_{inf}\times f_{inf}$, where $\eta_{MAC}$ denotes the number of MAC operations per inference cycle (1 MAC = 1 MUL + 1 ADD = 2 OPS), $B_{D}$ and $B_{W}$ represent the number of input and weight bits processed per MAC operation, respectively, and $f_{inf}$ is the inference frequency. The CIM macro is evaluated at $f_{inf} = 1 MHz$ frequency and achieves a peak throughput of 113 1b-GOPS, with an energy efficiency of 6.65 1b-TOPS/W. When considering the full system—including the input generation, weight updates, and output reading via the RISC-V core—the system reaches a peak throughput of 3.05 1b-GOPS, with an energy efficiency of 0.122 1b-TOPS/W.

\section{Conclusion}\label{sec:Concl}
\noindent Fabricating accurate and reliable  mixed-signal CIM architectures is still linked to several challenges, in particular the computation density, the integration in complete systems and the reliability of analog computing devices. In this work, we propose to use our proof-of-concept AI accelerator system Acore-CIM to demonstrate how mixed-signal CIM cores can be focused on these integration and reliability challenges. By combining contributions in the CIM core architecture (with MDAC weight cell), in the system integration (in a complete RISC-V based system and its open-source framework) and in full on-chip calibration (automatized by the RISC-V core), we manage to significantly improve the compute SNR of the system to achieve an average of 22 dB for each column, minimizing the accuracy loss due to CIM computation. With this work, we hope to open the path for reliability considerations in the design of mixed-signal CIM cores towards the development of end-to-end AI accelerators.  

\bibliographystyle{IEEEtran}

\bibliography{references}

\begin{IEEEbiography}[{\includegraphics[width=1in,height=1.25in,clip,keepaspectratio]{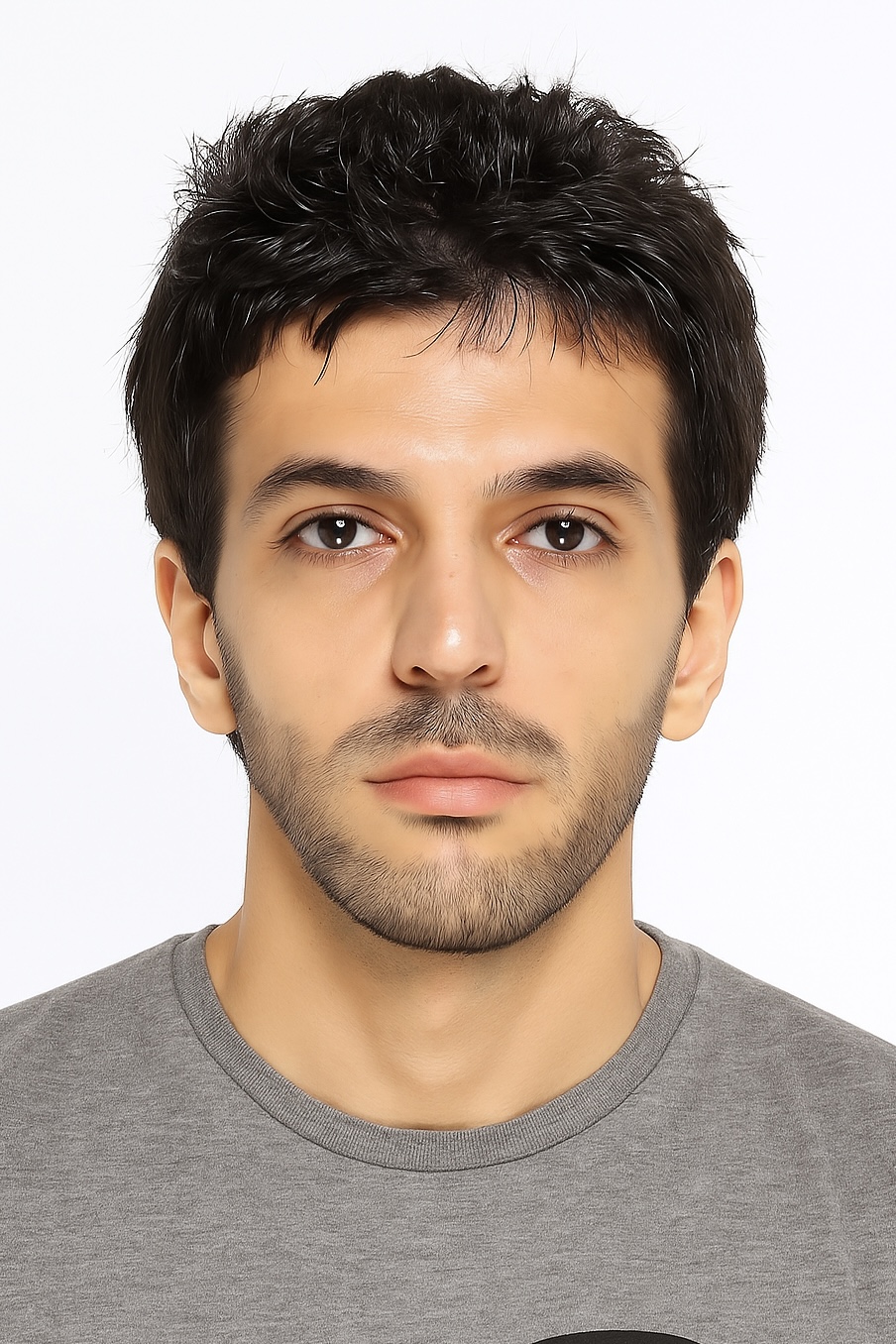}}]{Omar Numan (Student Member, IEEE)} received his M.Sc. degree in Micro- and Nano-electronic Circuit Design from Aalto University, Finland, in 2020, and is currently pursuing a Ph.D. at the same institution. His research focuses on designing advanced analog and mixed-signal integrated circuits, with a particular emphasis on compute-in-memory architectures for energy-efficient AI accelerators. He has contributed to the development of low-power, high-performance circuit designs that integrate memory and processing for next-generation AI hardware. 
\end{IEEEbiography}

\begin{IEEEbiography}[{\includegraphics[width=1in,height=1.25in,clip,keepaspectratio]{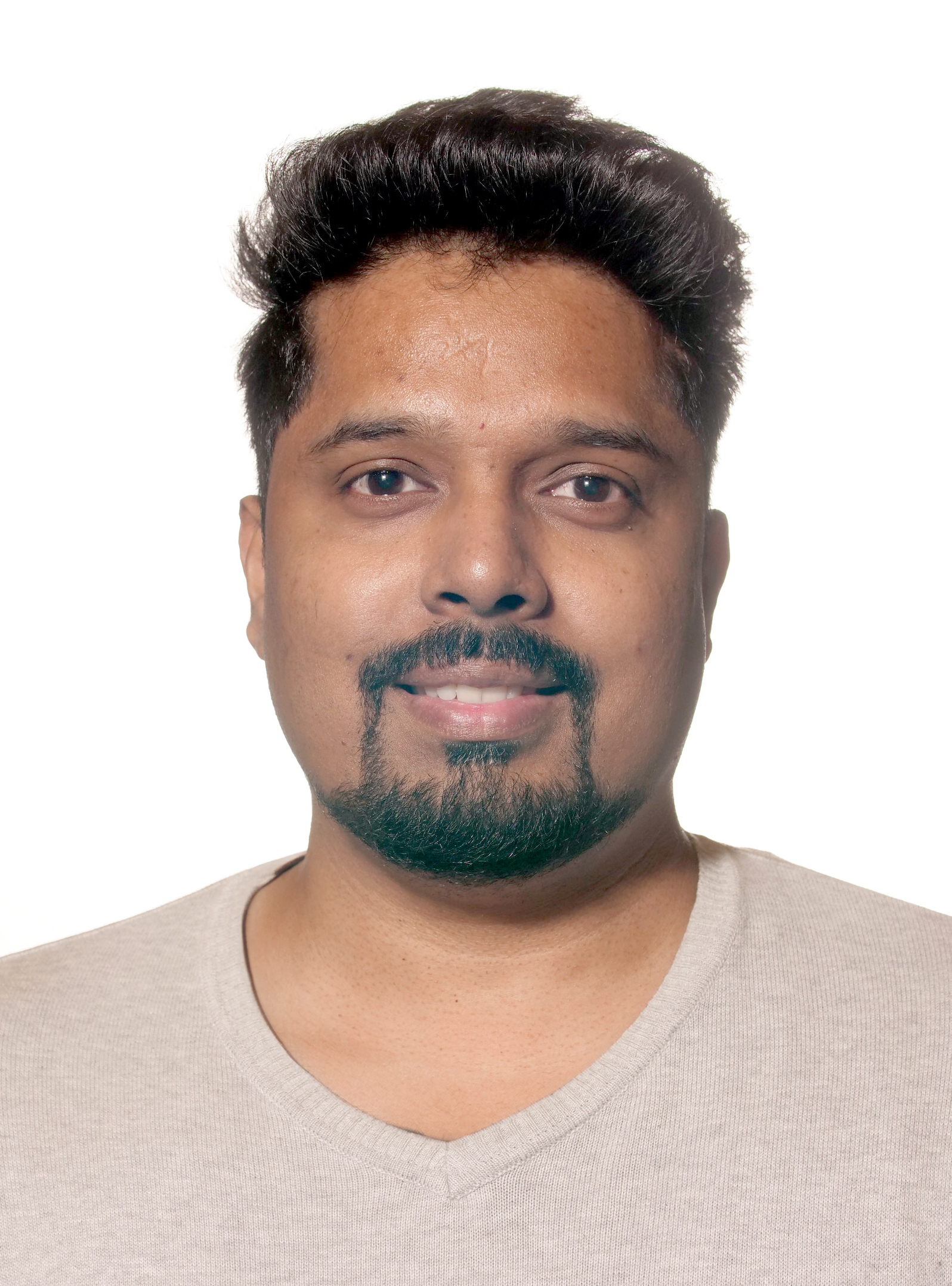}}]{Gaurav Singh (Student Member, IEEE)} earned his M.Sc. degree in VLSI from Amity University, Noida, India, in 2015. He is currently advancing his doctoral studies at the Department of Electronics and Nanoengineering, Aalto University, Finland. His research focuses on the development of low-power system-on-chip (SoC) solutions for sensor data processing. His work particularly explores microprocessors, low-power SRAM memories, and serial interfaces to enhance communication and debugging. His research work also includes integrating RISC-V processors with Compute In-Memory cores as hardware accelerators, aimed at improving power efficiency in AI applications.
\end{IEEEbiography}

\begin{IEEEbiography}[{\includegraphics[width=1in,height=1.25in,clip,keepaspectratio]{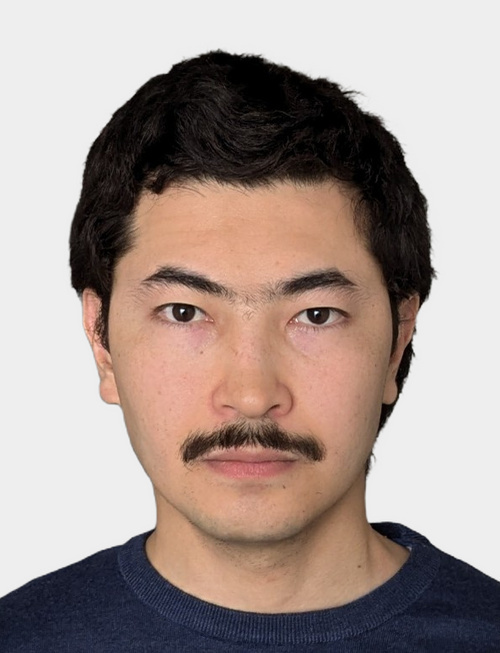}}]
{Kazybek Adam (Student Member, IEEE)} received his M.Sc. degree in Electrical and Electronic Engineering from Nazarbayev University, Kazakhstan, in 2019. He is currently pursuing his doctoral studies at the Department of Electronics and Nanoengineering, Aalto University, Finland. His research is in analog circuit design for AI accelerators with focus on Sample and Hold circuits, analog memories, and noise analysis.
\end{IEEEbiography}

\begin{IEEEbiography}[{\includegraphics[width=1in,height=1.25in,clip,keepaspectratio]{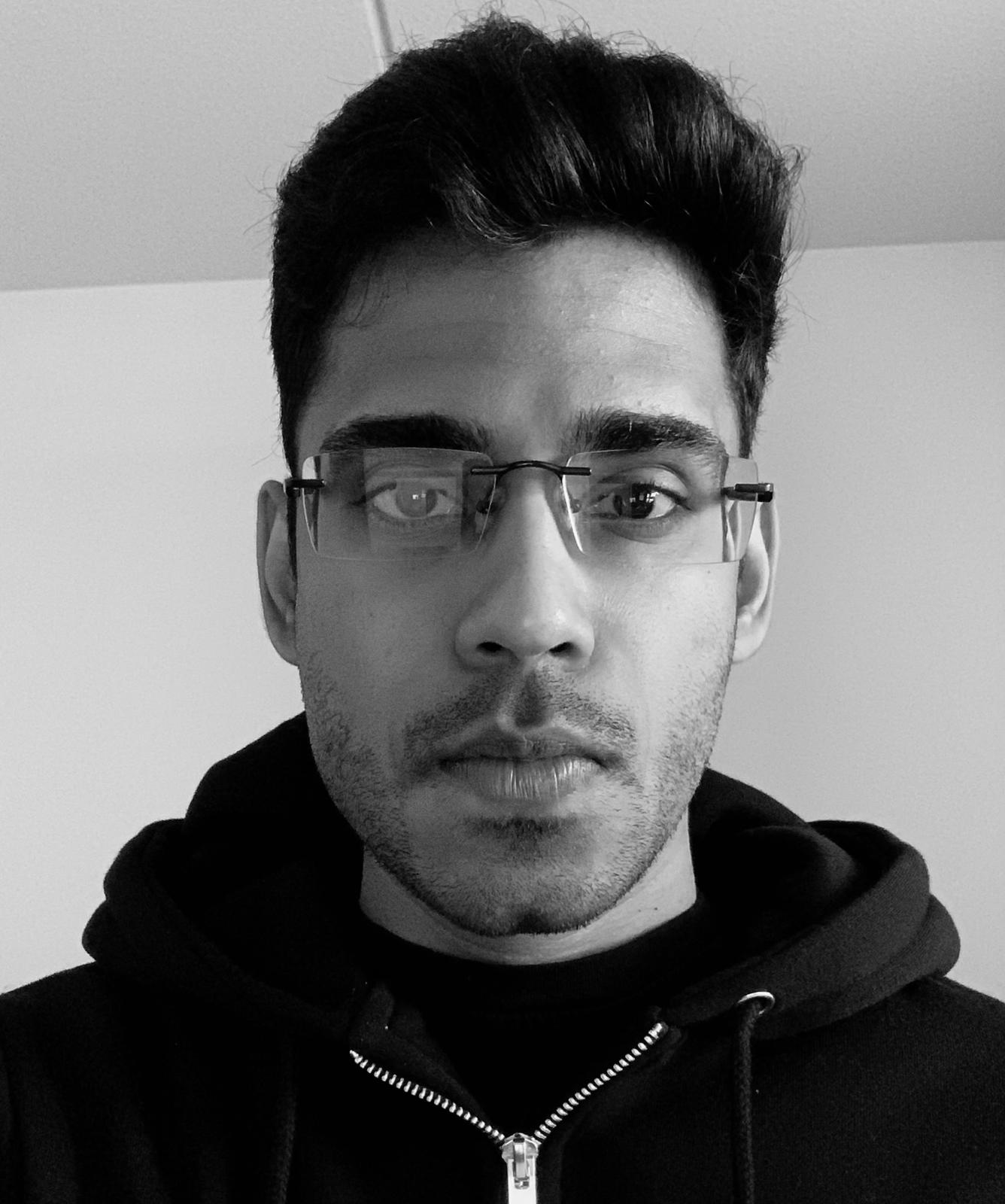}}]{Jelin Leslin (Student Member, IEEE)} 
is a doctoral candidate at Aalto University specializing in hardware-aware AI models. He earned a master’s degree in Electrical Engineering in 2020 from the Technical University of Eindhoven and KTH Royal Institute of Technology. His research focuses on energy-efficient AI model implementations through model compression, hardware-aware training, and custom hardware design. During his master's, he worked with Volvo Trucks on hardware acceleration for motion control in autonomous vehicles, which was also the focus of his thesis.
\end{IEEEbiography}

\begin{IEEEbiography}[{\includegraphics[width=1in,height=1.25in,clip,keepaspectratio]{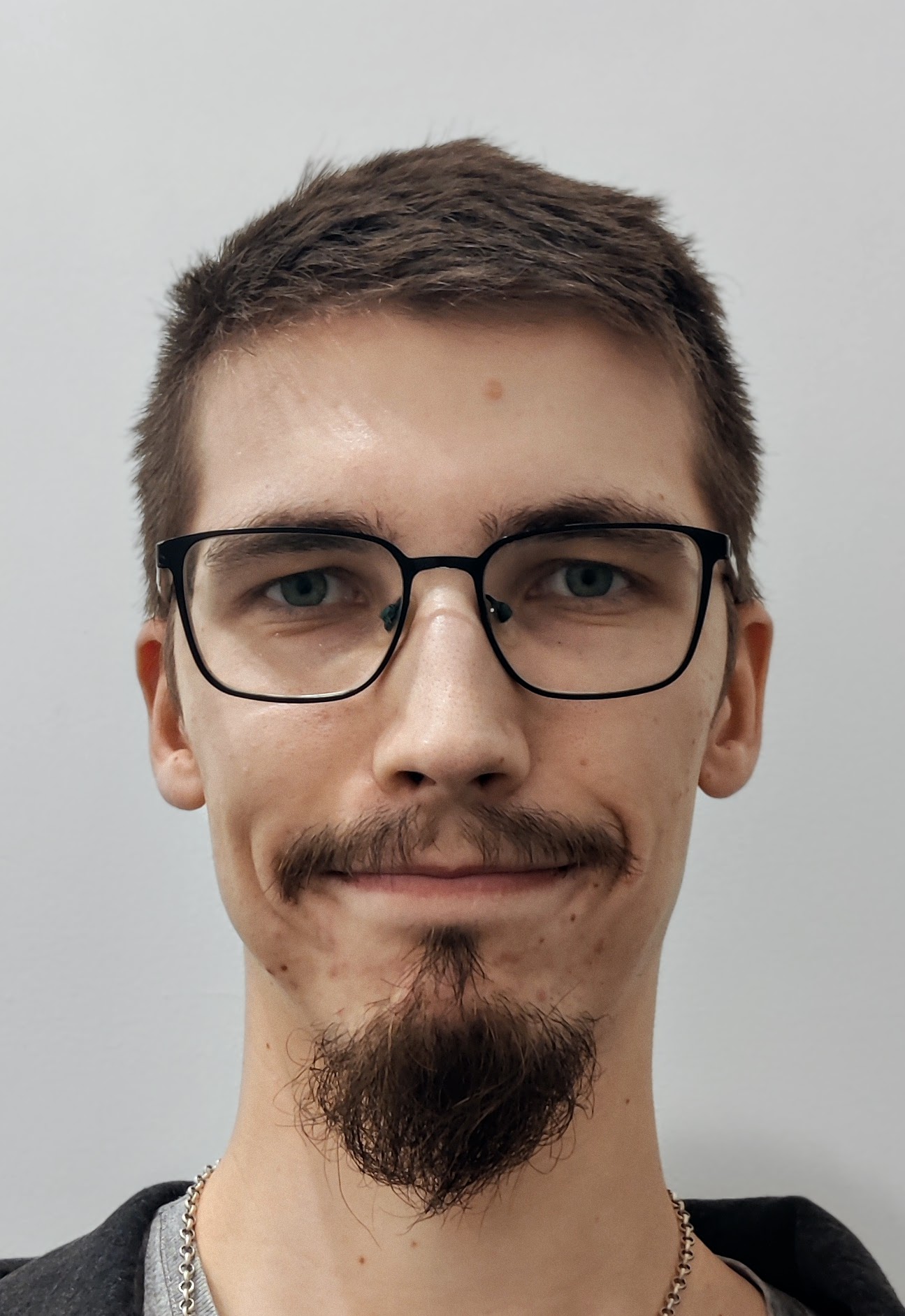}}]{Aleksi Korsman (Student Member, IEEE)} 
received his M.Sc. degree in Micro- and Nanoelectronic Circuit Design from Aalto University, Finland, in 2022. He is currently pursuing a Ph.D. degree at the same institution. His research interests include integrating hardware accelerators with microprocessors, especially for signal processing applications. 
\end{IEEEbiography}

\begin{IEEEbiography}[{\includegraphics[width=1in,height=1.25in,clip,keepaspectratio]{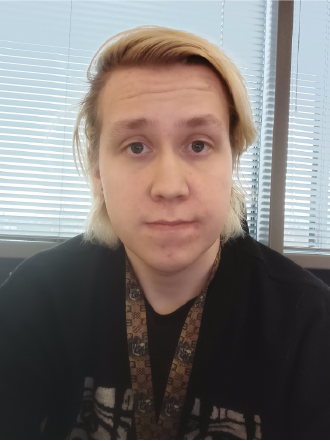}}]{Otto Simola (Student Member, IEEE)} 
received his M.Sc. degree in Micro- and Nanoelectronics Circuit Design from Aalto University, Finland, in 2023. Currently, he is pursuing he's Ph.D. degree at the institution. His research topic is programmatic circuit design especially relating to processor hardware acceleration and digital signal processing applications. 
\end{IEEEbiography}

\begin{IEEEbiography}[{\includegraphics[width=1in,height=1.25in,clip,keepaspectratio]{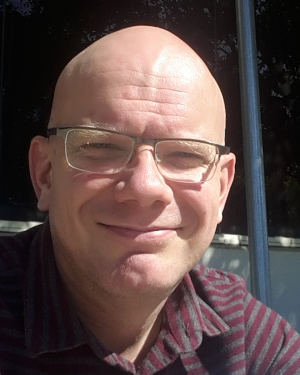}}]{Marko Kosunen (S\textquoteright97$-$M'\textquoteright07)} received his M.Sc, L.Sc and D.Sc (with honors) degrees from Helsinki University of Technology, Espoo, Finland, in 1998, 2001 and 2006, respectively. He is currently an Associate Professor at Aalto University, Department of Electronics and Nanoengineering. Academic years 2017-2019 he visited Berkeley Wireless Reserarch Center, UC Berkeley, on Marie Sklodowska-Curie grant from European Union. In addition to his academic duties, currently he is one of the three co-chairs of Microelectronics Finland, a academia-industry collaboration organization for microelectronics research and education. He has authored and co-authored more than hundred journal and conference papers and holds several patents. His current research interests include programmatic circuit design methodologies, digital intensive time-based data converters and transceiver circuits, and RISC-V microprocessor implementations with DSP accelerators.
\end{IEEEbiography}

\begin{IEEEbiography}[{\includegraphics[width=1in,height=1.25in,clip,keepaspectratio]{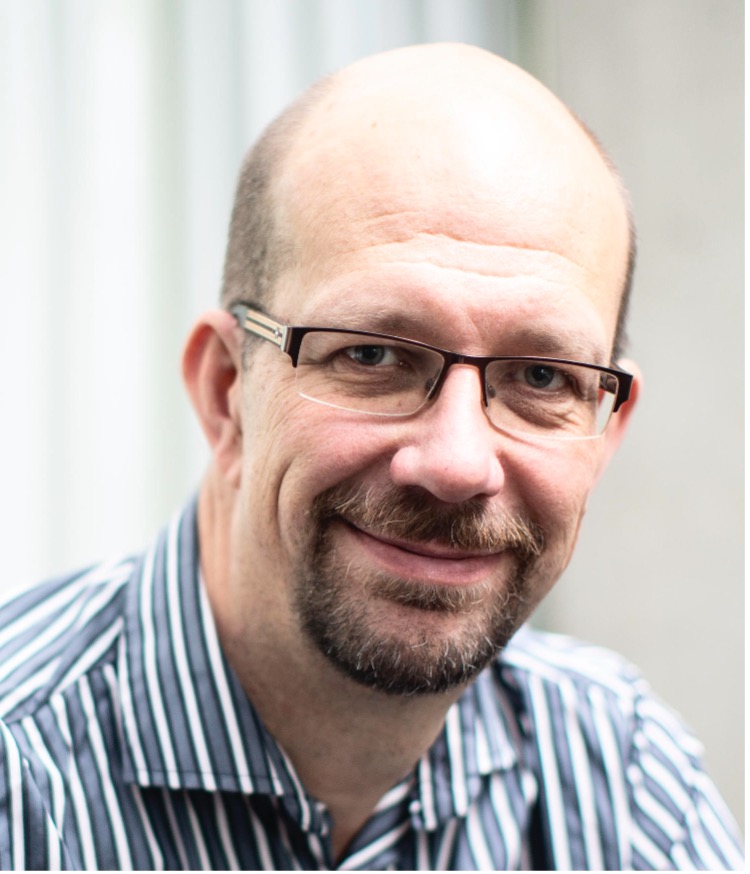}}]{Jussi Ryynänen (S’99$-$M’04$-$SM’16)} was born in Ilmajoki, Finland, in 1973. He received the M.Sc. and D.Sc. degrees in electrical engineering from the Helsinki University of Technology, Espoo, Finland, in 1998 and 2004, respectively. He is a full professor and the Dean of the School of Electrical Engineering, Aalto University, Espoo. He has authored or co-authored more than 200 refereed journal and conference papers in analog and RF circuit design. He holds seven patents on RF circuits. His research interests are integrated transceiver circuits for wireless applications. Prof. Ryynänen is currently an SG Member for the European Solid-State Circuits Conference (ESSCIRC) and the IEEE Nordic Circuits and Systems Conference (NORCAS). He has served as a TPC member of the IEEE International Solid-State Circuits Conference (ISSCC) and as a Guest Editor for the IEEE Journal of Solid-State Circuits.
\end{IEEEbiography}

\begin{IEEEbiography}[{\includegraphics[width=1in,height=1.25in,clip,keepaspectratio]{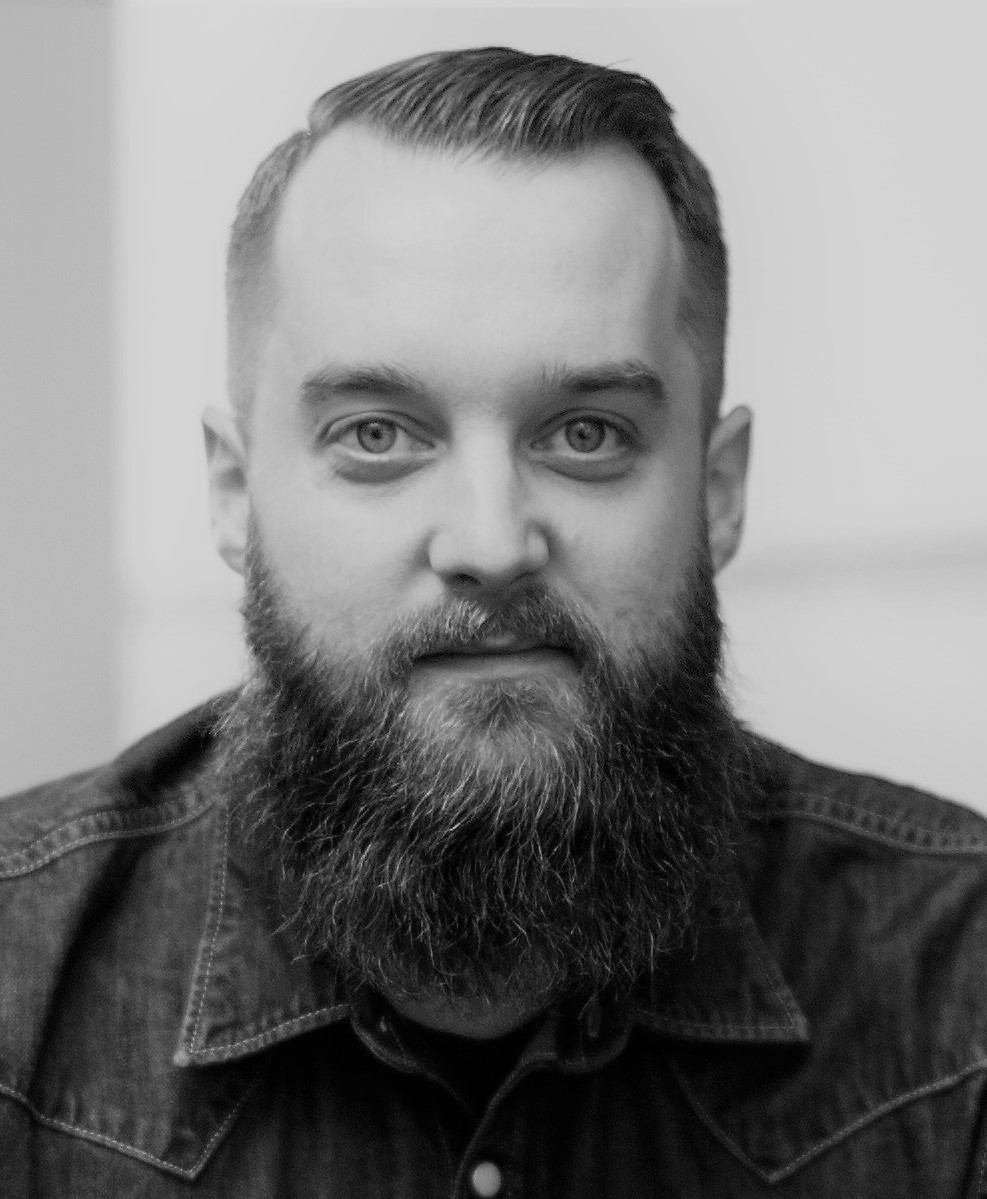}}]{Martin Andraud (Member, IEEE)} is an assistant professor in microelectronics at UCLouvain, Belgium, and a visiting professor at Aalto University, Finland. He received his PhD from Grenoble University, France, in 2016. He was a postdoctoral researcher successively with TU Eindhoven in 2016 and KU Leuven from 2017 to 2019. Between 2019 and 2024, he was an assistant professor at Aalto University, Finland. His research interests include the interface between edge AI, hardware/software co-design, testing, and reliability of custom ASIC for various AI accelerators, for instance, mixed-signal Compute-In-Memory architectures and alternative to deep learning models (probabilistic reasoning or hybrid AI). 
\end{IEEEbiography}

\vfill

\end{document}